\documentclass[%
    reprint,
    aps,
    superscriptaddress,
    showpacs,
    prb,
    floatfix,
    bibnotes,
    amssymb,
    amsfonts,
]{revtex4-2}

\usepackage[utf8]{inputenc}
\usepackage[T1]{fontenc}
\usepackage{amsmath,amsthm,mathtools,mathrsfs,physics}
\usepackage{graphicx}
\usepackage{subfig}
\usepackage{hyperref}
\usepackage{color}
\usepackage{siunitx}

\graphicspath{{img/}}

\DeclareSIUnit\angstrom{\text {Å}}

\begin{document}

\title{Superconducting transition temperatures of pure vanadium and vanadium-titanium alloys in the presence of dynamical electronic correlations}
\author{D. Jones}
\affiliation{Theoretische Physik III, Center for Electronic Correlations and Magnetism, Institute of Physics, University of Augsburg, 86135 Augsburg, Germany}
\affiliation{Augsburg Center for Innovative Technologies (ACIT), University of Augsburg, 86135 Augsburg, Germany}
\author{A. \"Ostlin}
\affiliation{Theoretische Physik III, Center for Electronic Correlations and Magnetism, Institute of Physics, University of Augsburg, 86135 Augsburg, Germany}
\author{A. Weh}
\affiliation{Theoretische Physik III, Center for Electronic Correlations and Magnetism, Institute of Physics, University of Augsburg, 86135 Augsburg, Germany}
\author{F. Beiu\c seanu}
\affiliation{Faculty of Science, University of Oradea, 410087 Oradea, Romania}
\author{U. Eckern}
\affiliation{Theoretische Physik II, Institute of Physics, University of Augsburg, 86135 Augsburg, Germany}
\author{L. Vitos}
\affiliation{Applied Materials Physics, Department of Materials Science and Engineering, Royal Institute of Technology, Brinellv\"agen 23, 10044, Stockholm, Sweden}
\author{L. Chioncel}
\affiliation{Theoretische Physik III, Center for Electronic Correlations and Magnetism, Institute of Physics, University of Augsburg, 86135 Augsburg, Germany}
\affiliation{Augsburg Center for Innovative Technologies (ACIT), University of Augsburg, 86135 Augsburg, Germany}

\date{\today}

\begin{abstract}
Ordinary superconductors are widely assumed insensitive to small concentrations of random nonmagnetic impurities, whereas strong disorder suppresses superconductivity, ultimately leading to a superconductor-insulator transition.
In between these limiting cases, a most fascinating regime may emerge where disorder enhances superconductivity. 
This effect is discussed here for the $\beta$-phase of vanadium-titanium alloys. Disorder is modeled using the coherent potential approximation while local electronic interactions are treated using dynamical mean-field theory. The McMillan formula is employed to estimate the superconducting transition temperature, showing a maximum at a Ti concentration of around $0.33$ for a local Coulomb interaction $U$ in the range of $\SIrange{2}{3}{\electronvolt}$.
Our calculations quantitatively agree with the  experimentally observed concentration dependent increase of $T_c$, and its maximal value of about $20\%$.
%
%
\end{abstract}

\maketitle

\section{Introduction}

The superconducting critical temperature $T_c$ is a key quantity connecting materials’ chemistry with the crystal and electronic structure of superconducting materials. 
At the same time $T_c$ emerges from the underlying mechanism of superconductivity.  
It is generally understood that superconductivity of many elemental metals at ambient pressures can be described by the Bardeen-Cooper-Schrieffer (BCS) theory~\cite{ba.co.57,ba.co.57b} based on the model of an effective instantaneous attractive interaction between electrons mediated by phonons (weak coupling limit). The BCS model was extended by considering the retarded nature of the electron-phonon interaction: in particular, the (strong) Coulomb repulsion between the paired electrons is reduced because of the large difference in electron and phonon velocities~\cite{mo.an.62}.
The proper retardation effects, beyond BCS, lead to a complex and frequency dependent order parameter, and is captured within the Eliashberg theory~\cite{elia.60}. 

The estimation of the critical temperature within Eliashberg theory involves moments of the Eliashberg function $\alpha^2 F(\nu)$, which describes the phononic modes through which electrons effectively interact, and their coupling to electrons.
The Eliashberg theory also accounts for the renormalization of the direct Coulomb interaction between two electrons ($\mu^*$) which is known as the pseudo-potential effect~\cite{mo.an.62}.
The McMillan formula allows the estimation of the critical temperatures from basic experimental data~\cite{mcmi.68}, including the electron mass enhancement factor due to the electron-phonon interaction, $m^*/m \approx 1+\lambda$. 

The alloys of transition metals constitute a wide class of superconducting materials. 
The theory of disordered superconductors was formulated within the seminal work of Abrikosov and Gorkov~\cite{ab.go.59}.
At the same time, Anderson~\cite{ande.59} has shown that superconductivity with $s$-wave pairing symmetry is insensitive to weak nonmagnetic disorder. 
Subsequent experimental studies, however, demonstrated that superconductivity is suppressed in strongly disordered samples. 
Very strong disorder can lead to a metal–insulator transition in the normal state, to the appearance of a pseudogap in the spectrum, to larger 
spatial fluctuations of superconductive pairing, and to an increased ratio of zero-temperature gap versus critical temperature $\Delta/T_c$~\cite{sh.ov.92,sado.97,gh.ra.98,gh.ra.01,st.br.08,sa.ch.08,ga.do.10} in agreement with the experimental results.

Thus, introducing random impurities and defects in ordered materials can be regarded as a tool for controlling the superconducting characteristics of materials. 
The main problem, however, is to identify the nature of the impurity distribution, or more exactly the shape of the probability distribution function for the disorder realizations.
In many materials simulations, the Coherent Potential Approximation (CPA)~\cite{sove.67,ve.ki.68,el.kr.74} is used. 
As the CPA is a local theory, it captures only the average presence of different atomic species, and therefore cannot account for more subtle aspects like possible short-range order.
These effects can be partly addressed using the dynamical cluster approximation (DCA)~\cite{ja.kr.01}, a cluster extension of dynamical mean-field theory (DMFT)~\cite{ge.ko.96,ko.vo.04,held.07} including electronic interactions, and the typical medium theory (TMT)~\cite{do.pa.03,dobr.10}. 
Some of these advances have been used for models~\cite{te.zh.17, we.zh.21} or in conjunction with material specific computations~\cite{mi.eb.17,os.zh.20} in the framework of density functional theory (DFT). In this context, we have recently compared different modeling approaches to disorder and estimated the possible superconducting temperature of the high entropy HfNbTaTiZr alloy~\cite{ka.mo.22}.

In the present paper we calculate the superconducting transition temperatures of pure vanadium (V) and vanadium-titanium (V-Ti) alloys, respectively, using the McMillan formula~\cite{mcmi.68}. 
We start by probing its capability in the case of the pure element V for which the description of dynamic electronic correlations is necessary to describe the Fermi surface in its normal state~\cite{we.be.17}. Further we continue in addressing the problem of the enhancement of the critical temperature in some of the V-Ti alloys. 
In Sec.~\ref{sec:methods} we provide a brief description of the computational methods: the LDA+DMFT framework using a muffin-tin based approach, followed by a brief presentation of the essentials of the McMillan formula to estimate the critical temperatures. Then, Sec.~\ref{sec:results}, an analysis of the electronic structure of vanadium and the vanadium-titanium alloys in the normal state subject to dynamical electronic correlations and the corresponding estimates for the critical temperatures is presented. We summarize our results in Sec.~\ref{sec:summary}.

\section{Methods}
\label{sec:methods}
Density functional theory methods incorporating dynamical electronic correlations have proven essential in modeling physical properties of materials containing narrow-band electronic states~\cite{ko.vo.04,ko.sa.06,held.07}. An additional difficulty is brought about by the existence of structural disorder.  Green's function based DFT-methods are able to address the combined dynamic electronic correlation and disorder effects efficiently, since they compute directly one-particle Green's function that can be averaged according to the various disorder realizations~\cite{mi.eb.17,os.vi.17,os.vi.18,os.zh.20}. 

We start with a short description of the LDA+DMFT method used in the present computations, and then describe the methodology for computing the Hopfield parameters~\cite{hopf.69} as formulated by Gaspary-Gyorffy in the language of the multiple scattering formalism~\cite{ga.gy.72}. 
This parameter contains the  electron-phonon coupling matrix elements which are essential for the estimation of the superconducting critical temperatures employing the McMillan formula~\cite{mcmi.68}, and can be directly used in combination with the interacting LDA+DMFT Green's function as implemented in standard packages of electronic structure calculation. 

\subsection{The combined disorder and dynamical electronic correlations}
\label{sec:DFT+DMFT}

To discuss dynamic correlation effects within the framework of LDA+DMFT~\cite{ko.sa.06,held.07} the standard methodology is to consider the following
multi-orbital on-site interaction term:
\begin{equation}
H_U = \frac{1}{2}
\sum_{i{m,\sigma}}
U_{mm^{\prime}
m^{\prime\prime}m^{\prime\prime\prime}} 
c^\dagger_{im\sigma} c^\dagger_{im^\prime \sigma^\prime} c_{im^{\prime\prime\prime}\sigma^{\prime}}
c_{im^{\prime\prime}\sigma}    
\end{equation}
Here $c_{im\sigma} (c^\dagger_{im \sigma})$ destroys (creates) an electron with spin $\sigma$ on orbital $m$ at the site $i$. 
The Coulomb matrix elements
$U_{mm^{\prime}m^{\prime\prime}m^{\prime\prime\prime}}$ are parameterized in terms of the average local Coulomb $U$ and exchange parameter $J$\cite{ko.sa.06}. 
In principle, the dynamical electron-electron interaction matrix elements can be computed~\cite{ar.im.04}, however substantial variations exists depending on the used local-orbitals~\cite{mi.ar.08}. 
We consider $U$ in the range from $\SIrange{0}{5}{\electronvolt}$ and $J = \SI{0.6}{\electronvolt}$.
We have checked that the results essentially do not change when increasing $J$ to, say, $\SI{0.9}{\electronvolt}$.

The LDA+DMFT implementation used here is a charge and self-energy self-consistent scheme extending the exact muffin-tin orbitals (EMTO) method~\cite{an.je.94.2,vi.sk.00,vito.01} using the local-density approximation (LDA).
The full d-manifold of states is considered in the many-body computations, while the charge self-consistency---through hybridization---ensures that dynamic correlation effects are felt by all orbitals.  
The impurity solver produces the many-body self-energy $\Sigma_\sigma(E)$ by the method of spin-polarized $T$-matrix fluctuation exchange~\cite{li.ka.97,po.ka.06,ir.ka.08}, and corrects the starting LDA Green's function as discussed in previous publications~\cite{ch.vi.03,os.vi.17,os.vi.18}. 
To eliminate double counting of the interactions already included
in the LDA exchange-correlation functional, the self-energy $\Sigma_\sigma(E)$ is replaced by $\Sigma_\sigma(E)-\Sigma_\sigma(0)$ in all equations of the LDA+DMFT scheme. 
Throughout this paper finite temperatures are only considered for the electronic
subsystem, where the temperature enters in the Matsubara frequencies $\omega_n = (2n + 1)\pi T$.
We perform total energy calculations within the charge self-consistent LDA+DMFT framework which follows the prescription in the literature~\cite{ko.sa.06}. The total energy is obtained from the LDA+DMFT functional~\cite{ko.sa.06} using the converged charge density and local Green’s function. 
In addition to the standard LDA total energy, the Kohn-Sham band energy correction due to the DMFT is added together with the trace of the matrix product between the DMFT self-energy and the  local Green's function ($\Sigma G$), the so-called Galitskii-Migdal contribution. 
For every pair of ($U,J$) values, the LDA+DMFT total energy was minimized as a function of the lattice parameter, and the superconducting critical temperature was computed.

Disorder was modeled through the CPA~\cite{sove.67,ve.ki.68} which allows computation for any fractional concentration of disorder realizations in multi-component alloys. 
Charge and self-energy self-consistency is achieved through the scattering path operator. This is the central quantity in the multiple scattering formulations of the electronic structure and allows the simultaneous computation of the real space charge densities through charge self-consistency and a proper normalization the Green's function (DMFT self-consistency).  
Technically the CPA implements an algebraic average of the scattering path-operator.
Details of the implementation and a comparison between CPA/TMT+DMFT methods have been published previously~\cite{os.zh.20,os.ch.21}. 

\subsection{McMillan critical temperatures from electronic structure computations}
\label{sec:Tc}
 
An estimate of the critical temperature in conventional superconductors can be obtained once the strength of the electron-phonon couplings is known.
There are significant advances in the first-principle calculations of $T_c$ using BCS-Eliashberg type of theories~\cite{ma.lu.05,go.am.09,no.gi.10}. However, since electron-phonon calculations in disordered alloys from first principles remain computationally expensive, simplified approaches like the Gaspari-Gyorffy theory~\cite{ga.gy.72} are used to provide the input for these estimates. 

Within the theory of strongly-coupled superconductors~\cite{mcmi.68} the electron-phonon coupling constant $\lambda$ can be
expressed as: 
\begin{align}
    \label{eq:lambda}
    \lambda=N(E_F) \expval*{I^2} / M \expval*{\omega^2}
\end{align}
where $N(E_F)$ is the density of states (DOS) at the Fermi level, $ \expval*{I^2}$ is the squared electron-phonon matrix element, averaged over the Fermi surface, $M$ is the
atomic mass, and $\expval*{\omega^2}$ is the average squared phonon frequency. 
The numerator, $\eta=N(E_F) \expval*{I^2}$, is known as the Hopfield parameter.
The observation that electron-ion interaction in transition metals has a $d$-resonance located in the vicinity of the Fermi surface allowed the formulation of a simple prescription to compute electron-phonon matrix-elements, and finally $\lambda$, within the multiple-scattering formalism~\cite{ga.gy.72}. Early studies attempted to estimate $\lambda$ for elemental metals from the knowledge of the experimental measured $T_c$ and the phonon spectra obtained from neutron scattering experiments~\cite{mcmi.68}. 
It was observed that although the individual values of the DOS at the Fermi level and the averaged electron-phonon matrix elements change across the 3d-series, the product of these two quantities remains approximately constant for all bcc-transition metals~\cite{hopf.69,ga.gy.72}. 
In a so-called local-phonon representation, the electron-phonon interactions mainly consist of scatterings that
change the electronic angular momentum $l$~\cite{hopf.69}.

Using these ingredients, Gaspari and Gyorffy~\cite{ga.gy.72}
proposed an approximate way to compute 
$\expval*{I^2}$ using the multiple-scattering Green’s function formalism and adopting the
rigid muffin-tin approximation.
In these approximations, the Hopfield parameter is computed from a combination of electronic scattering phase shifts and the electronic densities of states. 
The average squared phonon frequency $\expval*{\omega^2}$ 
can be approximated by using the Debye temperature $\theta_D$ via $\expval*{\omega^2} \approx \frac{1}{2} \theta^2_D$.
The same formalism was extended for disordered systems by including the proper disorder averaging of the physical quantities. Thus, the electron-phonon coupling constant for disordered alloys can be computed as $\lambda=\expval*{\eta}_d / \frac{1}{2} \expval*{M}_d \expval*{\theta_D}_d^2$, where $\expval{\dots}_d$ represents the disorder-averaged quantities.
As these parameters are accessible from a multiple-scattering based electronic structure calculation~\cite{go.gy.74}, the McMillan~\cite{mcmi.68} formula
\begin{equation}\label{eq:mcmilan_tc}
    T_c = \frac{\theta_D}{1.45} \exp[ -\frac{1.04(1+\lambda)}{\lambda-\mu^{*}(1+0.62\lambda)}],
\end{equation}
can then be directly applied. The factor $(1+\lambda)$ plays the role of an electron mass enhancement and the parameter $\mu^{*}$ is an effective Coulomb repulsion reflecting the retardation effect of electron-phonon coupling with respect to the instantaneous Coulomb repulsion~\cite{mcmi.68}. 
In our computations $\mu^*$  will be further considered as a variable parameter. All other ingredients of the formula besides $\mu^*$ are: $\lambda$, computed using the exact-muffin-tin-orbitals (EMTO) method~\cite{vi.sk.00,vito.01}, and the Debye temperature collected from experimental data.

\section{vanadium and vanadium-titanium alloys}
\label{sec:results}

The physical properties of V have been intensively studied~\cite{vo.br.11}. In fact, V is a type-II superconductor with a transition temperature of $T_c=\SI{5.38}{\kelvin}$. Its normal state thermodynamic properties were explained invoking phonons contribution~\cite{we.al.68}.

Electronic structure computations were also performed by various methods including standard LDA and extensions such as LDA+$U$~\cite{to.wa.03}, LDA+DMFT~\cite{we.be.17,be.ka.23} or including GW-like corrections~\cite{si.pa.20}. 
An alternative LDA+FLEX computation for V~\cite{sa.re.18} showed a sizeable, remarkably local ($\mathbf{k}$-independent) self-energy,
which demonstrates the importance of local dynamic correlation effects in computing the ground-state properties.
The LDA+DMFT treatment revealed significant quantum fluctuations, provided a better agreement with the experimental Fermi surface~\cite{we.be.17}, and allowed to formulate a microscopic theory for the transition from the Pauli paramagnetism to Curie-Weiss behavior~\cite{be.ka.23}.
Computed estimates for the critical temperature of V~\cite{lo.iy.03} under pressure, using the McMillan formula, were also reported to be in agreement with experiment.

For the Ti$_{x}$V$_{1-x}$ alloys, resistivity studies showed a relatively wide transition from the normal to the superconducting state which cannot be simply explained by the existence of a secondary superconductive phase~\cite{ma.sh.14}. 
The broad transition was observed for a relatively large concentration range.
Thus, proposals were formulated for additional scattering mechanism to the regular electron-phonon coupling based on spin-fluctuation theories~\cite{be.sc.66,ri.wi.79}. 
In addition, experiments showed that Ti doping in bcc-V leads to an enhanced $T_c$ which exceed the individual $T_c$ values of the alloy components. 
As dynamical electronic correlations seem to provide an improved picture of physical properties of pure V~\cite{we.be.17,si.pa.20,be.ka.23} it is natural to expect that these effects play also an important role for the disordered alloy. 
The aim, thus, is to explain the increase in the critical temperature, $T_c$, as a consequence of the combined effect of disorder and dynamical electronic correlations in the dilute concentration limit.

\subsection{Dynamic electronic correlation in the normal state of vanadium}
\label{sec:V}
In the following, we present the results of the electronic structure of pure vanadium computed within LDA+DMFT. 
Vanadium possess a body centered cubic (bcc) structure within  
the so-called $\beta$-phase. 
The space group corresponding to the $\beta$-phase is Im$\overline{3}$m with V occupying the Wyckoff $2a$ position, and the experimental lattice parameter is $a_{\mathrm{exp}}=\SI{3.02}{\angstrom}$.
To determine the equilibrium lattice parameter $a_{\mathrm{eq}}$ for the pure vanadium crystal, we computed the LDA+DMFT total energy of the system for different values of the interaction strength $U$ and a range of lattice parameters $a$ around the expected energy minimum using the formalism presented in Ref.~\cite{os.vi.17}. 
The convergence was checked on various $k$-mesh sizes up to $57 \times 57 \times 57$ $k$-points in the irreducible part of the first Brillouin zone, although the saturation of the results (no significant change in $N(E_F)$, or in the Hopfield parameters) was noticed for smaller number of $k$-points.

\begin{figure}[ht]
\includegraphics[width=0.99\linewidth,clip=true]{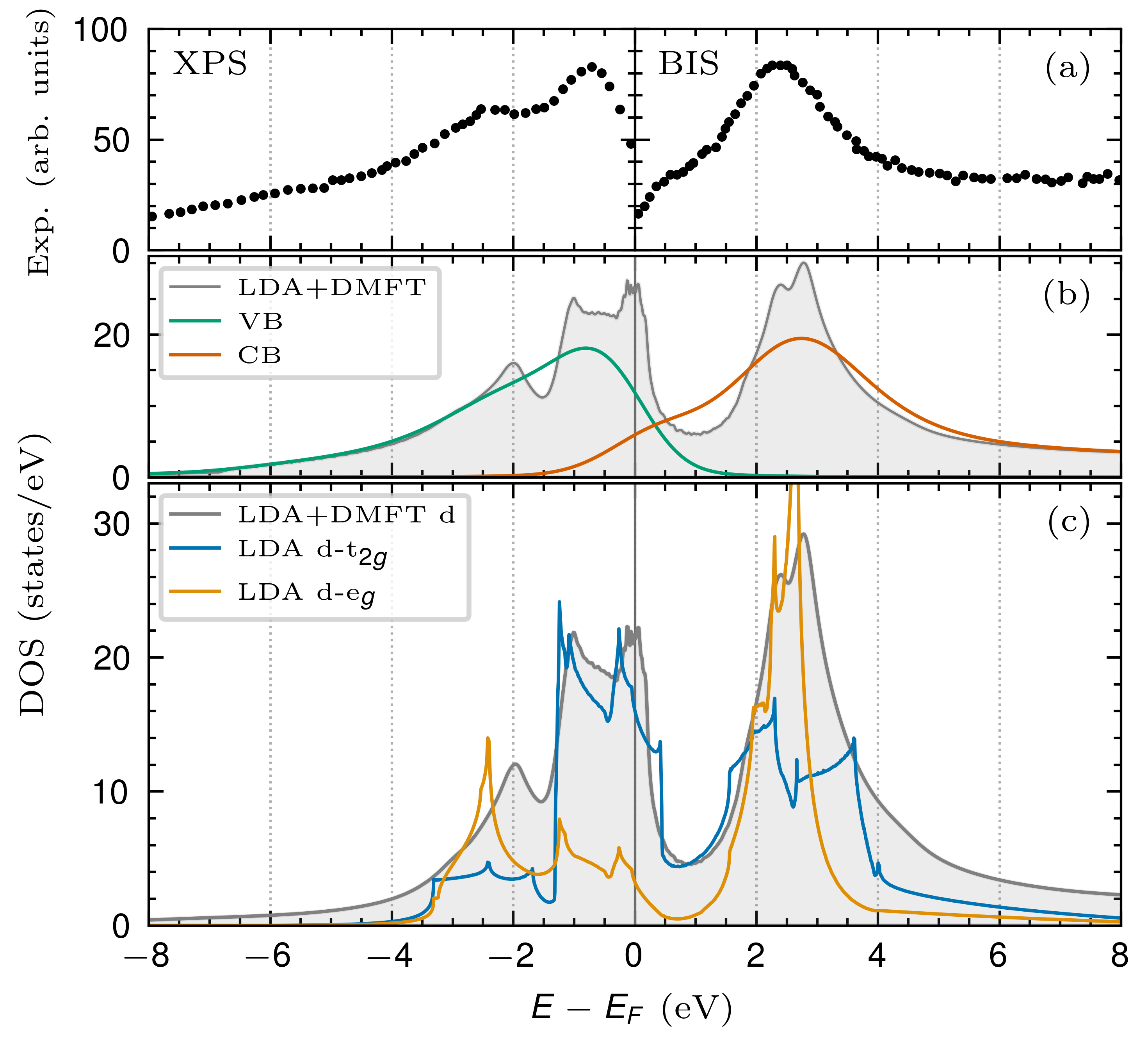}
\caption{Vanadium spectral functions. 
(a) Experimental XPS~\cite{le.da.77} and BIS~\cite{sp.fu.84} spectra in arbitrary units. 
(b) Valence band (VB) and conduction band (CB) DOS computed as convolution of the total density of states with distributions corresponding to the experimental resolution.
(c) DOS of the d-orbital of vanadium obtained using LDA and LDA+DMFT with $U=\SI{2}{\electronvolt}$ and $J=\SI{0.6}{\electronvolt}$ at \SI{300}{\kelvin}.
}\label{fig:DOS_V}
\end{figure}

The electronic structure calculations provide information about the electronic bands and the density of states. The computed results can be compared to experimental measurements with different spectroscopic techniques such as the X-ray photoemission spectra (XPS)~\cite{le.da.77} for the valence band, and Bremsstrahlung isochromat spectra (BIS)~\cite{sp.fu.84} for the conduction band DOS or the one-particle spectral function. 
The XPS spectra of the valence band are seen on the left side in the panel (a) of Fig.~\ref{fig:DOS_V}, while panel (b) show results of the empty electronic states up to $\SI{8}{\electronvolt}$ above $E_F$ using BIS~\cite{sp.fu.84}. 
Within the valence band (VB) the two main features are the broad peaks at energies at about $\SI{-2.4}{\electronvolt}$ and $\SI{-0.7}{\electronvolt}$. 
In the conduction band (CB) a hump at about $\SI{0.7}{\electronvolt}$ and a larger peak at $\SI{2.3}{\electronvolt}$ can be seen.
The computed total DOS obtained using LDA+DMFT can be seen in Fig.~\ref{fig:DOS_V}(b).
In order to compare the experimental and theoretical results for the VB the computed DOS is multiplied by the Fermi-Dirac distribution $f_{FD}(E,T)$ and for the CB the multiplication is performed with the factor $(1-f_{FD}(E,T))$.
In addition, the spectra of VB and CB are convoluted with a constant Lorentzian broadening of $\SI{0.1}{\electronvolt}$ followed by a Gaussian broadening of $\SI{0.55}{\electronvolt}$ ($\SI{0.7}{\electronvolt}$) corresponding to the experimental resolution~\cite{si.pa.20}. 
After the application of broadening and convolution functions to the full DOS the theoretical spectra are significantly smoothed out. Nevertheless, they capture the formation of the pseudo-gap above the Fermi energy $E_F$ and the position of the main peaks in the spectral function. 
Note also the presence of tails in the LDA+DMFT DOS at high binding energies ($E-E_F  \in [-8, -4] \si{\electronvolt}$) which cannot be captured in plain LDA as these are characteristic features of quantum fluctuations seen in many transition metal elements. A particular example is the formation of the so-called $-\SI{6}{\electronvolt}$ satellite in Ni~\cite{li.ka.01,br.mi.06,ko.po.12}.

Figure~\ref{fig:DOS_V}(c) shows a direct comparison of LDA and LDA+DMFT densities of states. 
The gray area corresponds to the LDA+DMFT total d-orbitals DOS while the blue and orange lines are the LDA V-d t$_{2g}$/e$_g$-orbitals DOS. 
Note that within LDA the d-electrons bandwidths of the t$_{2g}$ and e$_g$-orbitals have similar energy extension. 
At the Fermi level a dominant d-t$_{2g}$ contribution can be seen. The negative slope of the real part of the self-energy in the LDA+DMFT computation leads to the slight compression of spectra towards the Fermi level, which is a characteristic Fermi liquid behavior. This effect is easily visible as the position of the d-e$_g$-orbital peak in the vicinity of $\SI{-2.5}{\electronvolt}$ in the valence band is shifted closer to about $\SI{-2}{\electronvolt}$ in the LDA+DMFT results, and at the same time a more substantial DOS is obtained at the Fermi level. 
The unoccupied part dominated by the e$_g$-DOS peak at about $\SI{2.7}{\electronvolt}$ is slightly shifted in this case away from the Fermi level, which indicates that a mean-field Hartree-Fock like contribution dominates dynamic local correlations in this energy range.
Further discrepancies between the measurement and computed results in both occupied and unoccupied 
parts of the spectra have been previously addressed~\cite{si.pa.20}. It was concluded that alternative computations such as LDA+$U$ fail to describe the weight of computed spectra situated between the main peaks. It was also emphasized that electron-electron interactions describes by variants of GW and LDA+DMFT bring the computed spectra in fairly good agreement with the experiment~\cite{si.pa.20}.

\begin{figure}[ht]
\includegraphics[width=0.99\linewidth,clip=true]{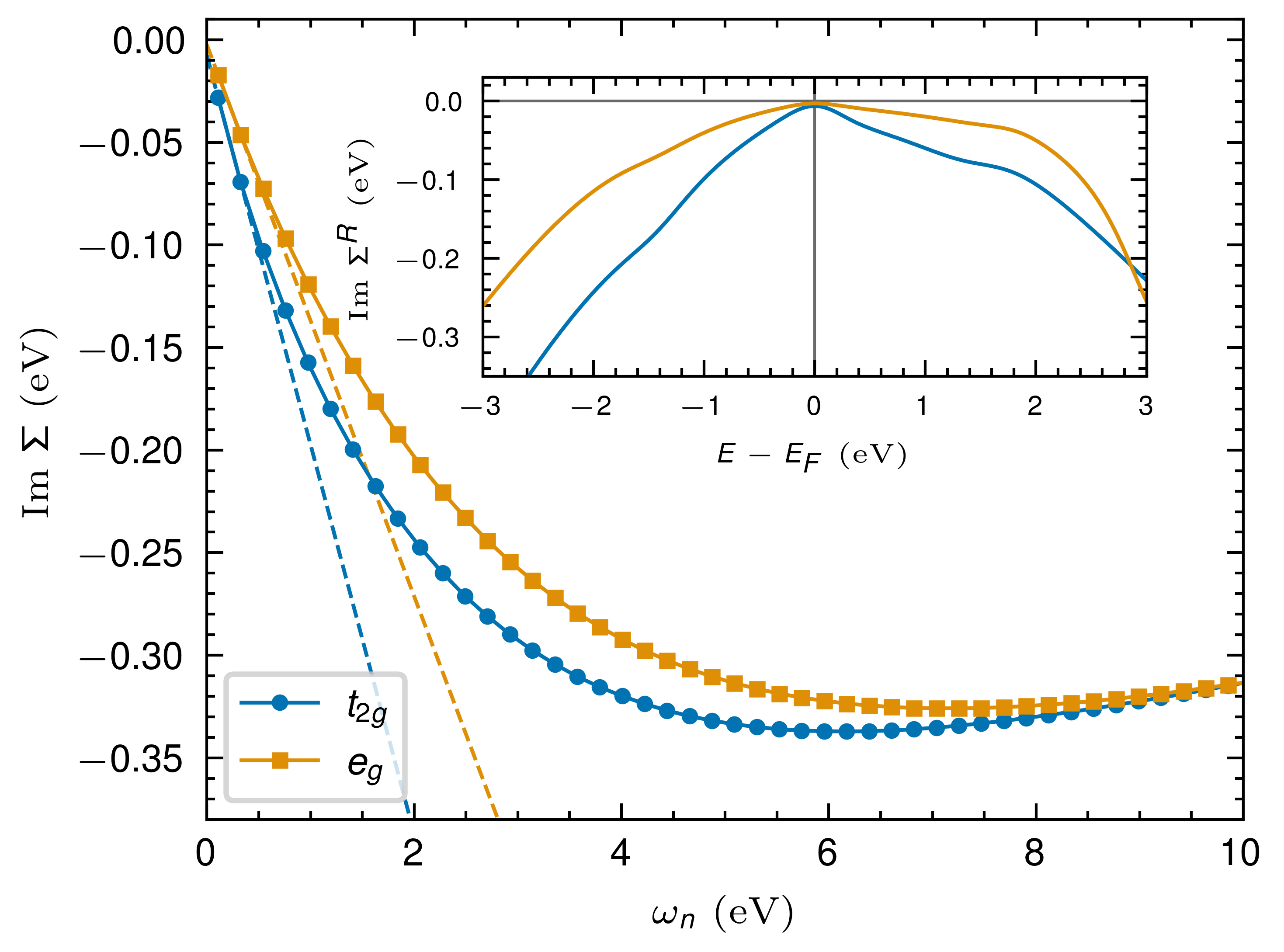}
\caption{Imaginary part of the self-energies as a function of Matsubara frequencies of the vanadium t$_{2g}$ and e$_g$ orbitals obtained from LDA+DMFT with $U=\SI{2}{\electronvolt}$ and $J=\SI{0.6}{\electronvolt}$ at \SI{400}{\kelvin}. The inset shows the parabolic behavior of the imaginary part of the retarded self-energy around $E_F$, indicating Fermi liquid behavior. 
}\label{fig:Sig_V}
\end{figure}

In the following, we discuss the effects of dynamical correlations based on the results of the electronic self-energies. 
In Fig.~\ref{fig:Sig_V} we show the imaginary part of the orbital resolved self-energies. 
The inset shows the analytically continued retarded self-energies on the real axis which follows a quadratic energy dependence in the vicinity of the Fermi level: $\Im \Sigma^R(E) \propto -(E-E_F)^2$. 
Both the t$_{2g}$/e$_g$-orbitals show a Fermi liquid behavior. However, as expected the effective masses are different. The analytically continued self-energies are obtained by using Pad\'e-approximate methods~\cite{vi.se.77,we.ot.20}. 
To avoid the possible inaccuracies of the analytical continuation, we interpret the computational results using the self-energy on the Matsubara axis.
The Fermi liquid state is characterized by a linear dependence of the imaginary part of the self-energy on imaginary frequencies $\omega_n$:   $\Im\Sigma(\omega_n)=-\Gamma(\omega_n)-(Z^{-1}-1)\omega_n$, where $Z$ is the quasiparticle spectral (inverse of the effective mass ratio $m/m^*$) weight and $\Gamma(\omega_n)$ the quasiparticle damping.
In the region of low Matsubara frequencies $\Im \Sigma_{t_{2g}/e_g}(\omega_n)$ approaches zero linearly (dashed lines Fig.~\ref{fig:Sig_V}) for both orbitals. 
Thus, the t$_{2g}$/e$_g$-electrons are long-lived quasiparticles, i.e. their scattering rate vanishes as the Fermi surface is approached. 
In the entire temperature range $\SI{100}{\kelvin} \le T \le \SI{600}{\kelvin}$ the linearity of the imaginary part of the self-energy  is preserved, thus the results provide a clear indication of the Fermi liquid character of the electronic system.

\begin{figure}[ht]
\includegraphics[width=0.99\linewidth,clip=true]{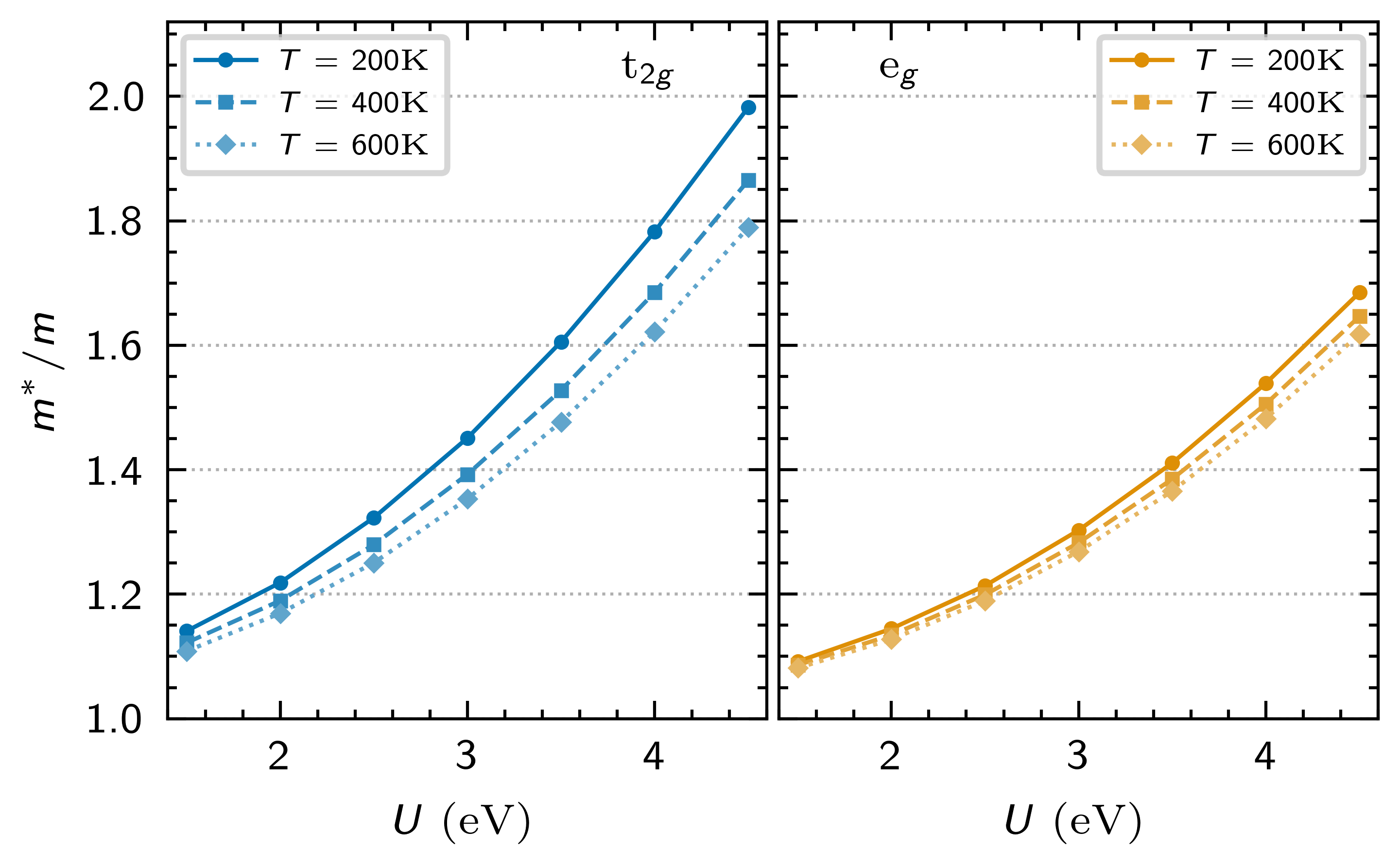}
\caption{Interaction strength dependence of the effective mass
renormalization of the t$_{2g}$ and e$_g$ orbitals of vanadium for different temperatures.
}\label{fig:meff_V}
\end{figure}

Within DMFT the quasiparticle mass enhancement $m^{\star}/m$, relative to the band mass $m$, becomes local and can be approximately computed from the Matsubara frequencies in the zero-temperature limit as $m^{\star}/m=1-\mathrm{Im}\Sigma(\omega_n)/\omega_n|_{\omega_n\rightarrow 0}$. 
It is often computed numerically in Quantum Monte-Carlo (QMC) calculations as the slope of the imaginary part of the self-energy for a few low Matsubara frequencies~\cite{ge.ko.96}. In Fig.~\ref{fig:Sig_V} dashed lines show the slopes for the t$_{2g}$ and e$_{g}$.
The results are presented in Fig.~\ref{fig:meff_V} for various Hubbard parameters up to $U=\SI{5}{\electronvolt}$. 
In computations with $U$ above $\SI{2}{\electronvolt}$ we used a constant $J=\SI{0.6}{\electronvolt}$, while for $U < \SI{2}{\electronvolt}$ we kept a constant $U/J = 2/0.6 \approx 3.33$ ratio to ensure the positiveness of the effective Coulomb repulsion $U-3J$.
Increasing the value of the Hund's coupling up to $J=\SI{0.9}{\electronvolt}$ no significant change in the spectral function occurs.
For all temperatures the effective mass of both orbitals is increasing with increasing $U$ values. 
At higher temperatures we estimate a somewhat smaller value. 
As seen from Fig.~\ref{fig:meff_V} the effective mass enhancements $m^*/m$ are in the range $1 \dots 2$, with an increasing tendency for larger $U$ values. 
Enhancements in this range are characteristic for medium-correlated electronic systems, in contrast to heavy fermion or strongly correlated systems for which $m^*/m$ can be of the order of hundred or thousand~\cite{pi.no.66,ne.or.88}. 
Thus our results indicate the presence of moderate dynamical correlation effects for both orbitals, but for the e$_g$ electrons the effect is less substantial. 

Deviations from the Fermi liquid behavior and appearance of non-analytic frequency dependence of the self-energy for the t$_{2g}$ states have been recently discussed in Ref.~\cite{be.ka.23}.  
Unlike our computation, which is based on an perturbative impurity solver, the strong dynamical correlations are more precisely captured by the continuous-time QMC solvers~\cite{gu.mi.11} used in Ref.~\cite{be.ka.23}. 
In our computations we did not address the high temperature and high $U$ limit.  
Nevertheless, for the relevant temperatures and $U$ range (between $2$ and \SI{3}{\electronvolt}) which correctly capture Fermi surface features~\cite{we.be.17} our results agree well with the QMC results~\cite{be.ka.23}, as well as with the LDA+FLEX calculations~\cite{sa.re.18} reported earlier.

\subsection{McMillan estimates for the superconducting transition temperature of pure V}
 
\textit{Ab-initio} calculations of superconducting temperatures have been attempted using the Eliashberg equation and produced prominent results~\cite{ma.lu.05,go.am.09,no.gi.10}. These computations did not include the presence of dynamic electronic correlations in disordered alloys.
However, the combined dynamical electronic correlation and disorder problem may be considered directly in the McMillan theory, which provides reliable estimates in such cases.  

\begin{figure}[ht]
\includegraphics[width=0.95\linewidth,clip=true]{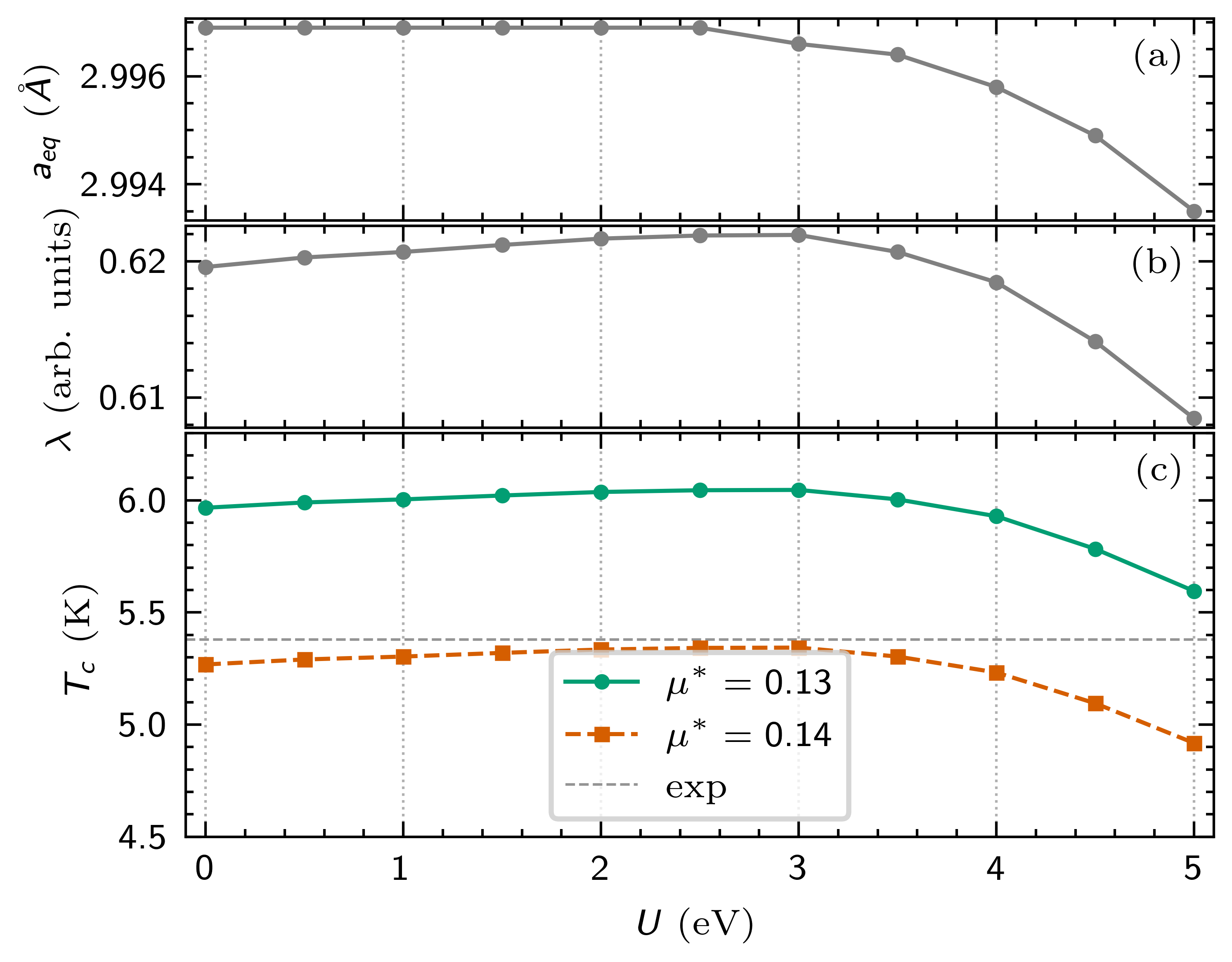}
\caption{(a) Computed equilibrium lattice parameter $a_{\mathrm{eq}}$,
(b) electron-phonon coupling strength $\lambda$, and (c) the superconducting critical temperature $T_c$ with respect to the interaction strength $U$  of vanadium. The dashed line indicates the experimental $T_c$. The experimental lattice parameter is $a_{\mathrm{exp}}=\SI{3.02}{\angstrom}$; it is not shown as it falls outside the shown scale.
}\label{fig:TcU_V}
\end{figure}

Figure~\ref{fig:TcU_V} summarizes the results of the LDA+DMFT computations: the equilibrium lattice parameter is seen in subplot (a), the mass-enhancement factor $\lambda$ is shown in panel (b) and finally panel (c) shows the variation of the superconducting temperature. Computations were performed for $U$ values up to $\SI{5}{\electronvolt}$ and for a constant $J=\SI{0.6}{\electronvolt}$. The values for $U = \SI{0}{\electronvolt}$ corresponds to the LDA results.
All quantities presented in Fig.~\ref{fig:TcU_V} have a slight variation with the strength of the Coulomb interaction $U$. The equilibrium lattice constant does not change significantly up to $\SI{3}{\electronvolt}$ and for larger $U$ values has a descending trend. For the value $U=\SI{5}{\electronvolt}$ a reduction of about $2\%$ is obtained. 
It is interesting to note that the mass enhancement factor $\lambda$ reaches a maximum value of $0.622$ for the realistic $U$ values in the range from $\SIrange{2}{3}{\electronvolt}$. For higher $U$ values $\lambda$ decreases as well.

With the first principles estimates $N(E_F), \lambda$ and the experimental data of Debye temperatures ($\theta_D$) the values of $T_c$ are obtained choosing a suitable value for $\mu^*$. McMillan suggested taking $\mu^*$ in the vicinity of $0.13$ for transition metal elements~\cite{mcmi.68}, however one finds different $\mu^*$ values in the current literature. Ref.~\cite{vora.10,ma.sh.14} observed a range of $\mu^*$ between $0.11$ to $0.14$ computed via various forms of local field correction functions~\cite{vora.10}, or have chosen values of $\mu^*$ between $0.12$ and $0.175$ to match their results to experimental data~\cite{ma.sh.14}.  
Fig.~\ref{fig:TcU_V}(c), shows our estimates for $T_c$: choosing a value of $\mu^*=0.13$ our results for pristine vanadium slightly overestimate the transition temperature, whereas for a value of $\mu^*=0.14$ a better match with experiment is found. 
Although the transition temperature $T_c$ is shifted for different choices of the effective Coulomb repulsion parameter $\mu^*$, the interaction strength dependence of $T_c$ shows the same trend:
a mild increase with $U$, a maximum value in the range of $U$ in which $\lambda$ is maximum, and a decreasing trend for higher $U$ values.  
Our overall results of LDA+DMFT computations with the Hubbard parameter values $U$ in the range of $\SIrange{2}{3}{\electronvolt}$ and $\mu^*=0.14$ provides $T_c$ next to the experimental value $T_c =\SI{5.73}{\kelvin}$, the dashed line of Fig.~\ref{fig:TcU_V}(c). Note also, that the same Hubbard $U$ values proved to provide an excellent description of Fermi surface of vanadium~\cite{we.be.17}.

\subsection{Binary Ti-V alloys}
\label{sec:V-Ti}
Ti$_x$V$_{1-x}$ alloys exist in different metallurgical phases. 
For concentrations $0 \leq x \leq 0.68$ the $\beta$ phase is observed with a bcc lattice structure. The $\alpha$ phase, is formed for high concentrations of Ti, $0.86 \leq x \leq 1$, in which the system has a hexagonal close-packed (hcp) structure. For concentrations in the range of $0.67 < x < 0.86$ an additional hexagonal $\omega$-phase (precipitation) can occur apart from the $\alpha$- and 
$\beta$-phases~\cite{ma.sh.14}.
The $\omega$-phase is a primitive hexagonal crystal structure that has been found to be a common metastable phase in body-centered cubic metals and alloys. In general, $\omega$-phase precipitates and has a coherent interfacial structure with its bcc matrix phase, with the parameters $a_\omega=\sqrt{2}a_{bcc}$ and $c_\omega=\sqrt{3}/2 a_{bcc}$.
In such a mixed phase an anomalous increase in electrical resistivity with decreasing temperatures was reported~\cite{sa.mu.90}.
This behavior was addressed invoking various models including weak localization~\cite{sa.mu.90}, Kondo ($s-d$)-interactions and localized spin-fluctuation~\cite{pr.ra.75a,pr.ra.75b}.
In addition, for concentrations $x \le 50\%$~\cite{ma.sh.14} an increase in the superconducting $T_c$ has been observed. As previously noted~\cite{ma.sh.14} this effect cannot be associated with the change in the Debye temperature as Ti and V atoms are situated aside in the periodic table. 
It is our aim to investigate whether dynamic electronic correlation effects contribute to the enhancement of the $T_c$. 

We consider in the following the bcc-$\beta$-phase in which Ti atoms occupy randomly the V $2a$ site with concentrations up to $65\%$.   
We have computed the total energy of the system
for different $U$ and $x$ values and determined the equilibrium lattice parameter $a_{\mathrm{eq}}$. After calculating multiple $a_{\mathrm{eq}}$, a linear concentration dependence of the equilibrium lattice parameter could be observed, which is also known as Vegard's law~\cite{de.as.91}:
$ a_{\mathrm{eq}} = x \cdot a_{\mathrm{eq}}^{\mathrm{Ti}} + (1 - x) \cdot a_{\mathrm{eq}}^{\mathrm{V}}$,
where $a_{\mathrm{eq}}^{\mathrm{V}}$ is the equilibrium lattice constant of pure vanadium and $a_{\mathrm{eq}}^{\mathrm{Ti}}$ is the (extrapolated) equilibrium lattice constant of pure titanium in the bcc phase.  
The extrapolated value of $a_{\mathrm{eq}}$ for a pure titanium crystal in the bcc-$\beta$-phase is $a_{\mathrm{eq}}^{\mathrm{Ti}} = \SI{3.231}{\angstrom}$. 

\begin{figure}[ht]
\includegraphics[width=0.95\linewidth,clip=true]{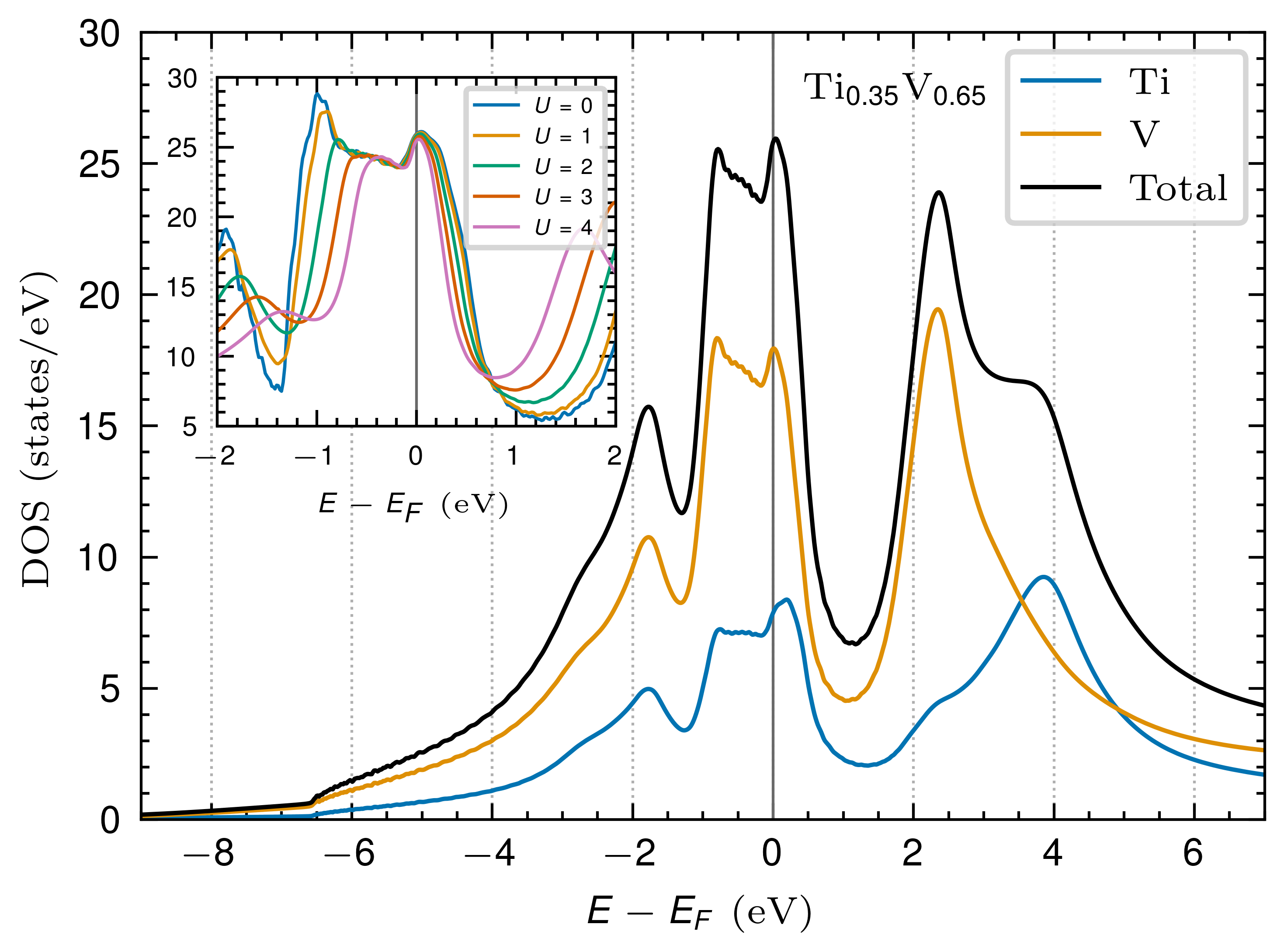}
\caption{Total and sublattice DOS of Ti$_{0.35}$V$_{0.65}$ obtained using LDA+DMFT with $U=\SI{2}{\electronvolt}$ and $J=\SI{0.6}{\electronvolt}$ at $\SI{400}{\kelvin}$.
The inset shows the total DOS for different interaction strengths $U$ around $E_F$.
}\label{fig:TiV_DOS}
\end{figure}

Figure~\ref{fig:TiV_DOS} shows the electronic DOS of the Ti$_x$V$_{1-x}$ alloy for $x=0.35$ concentration at which the superconducting critical temperature is the highest among all Ti$_x$V$_{1-x}$ alloys.
Independent of the impurity concentration $x$, the electronic structure is dominated by broad d-electrons bands with a specific pseudo-gap in the DOS as a consequence of the bcc-structure.  
As can be seen in Fig.~\ref{fig:TiV_DOS} most of the contribution at the Fermi level is provided by the V-d states. 
The inset shows a comparison of LDA+DMFT results for DOS for various $U$ values. Similarly to the computations for vanadium, we used a constant $J=\SI{0.6}{\electronvolt}$ for $U$ above $\SI{2}{\electronvolt}$, while for $U < \SI{2}{\electronvolt}$ we kept the ratio $U/J = 2/0.6 \approx 3.33$ constant.
We applied DMFT to both Ti and V components using the same Hubbard $U$ value.
The computation with $U=0$ corresponds the LDA-CPA calculation. One can see a clear tendency: a tiny reduction of DOS at the Fermi level and a narrowing of the d-bands in the vicinity of the Fermi level. 

\begin{figure}[ht]
\includegraphics[width=0.95\linewidth,clip=true]{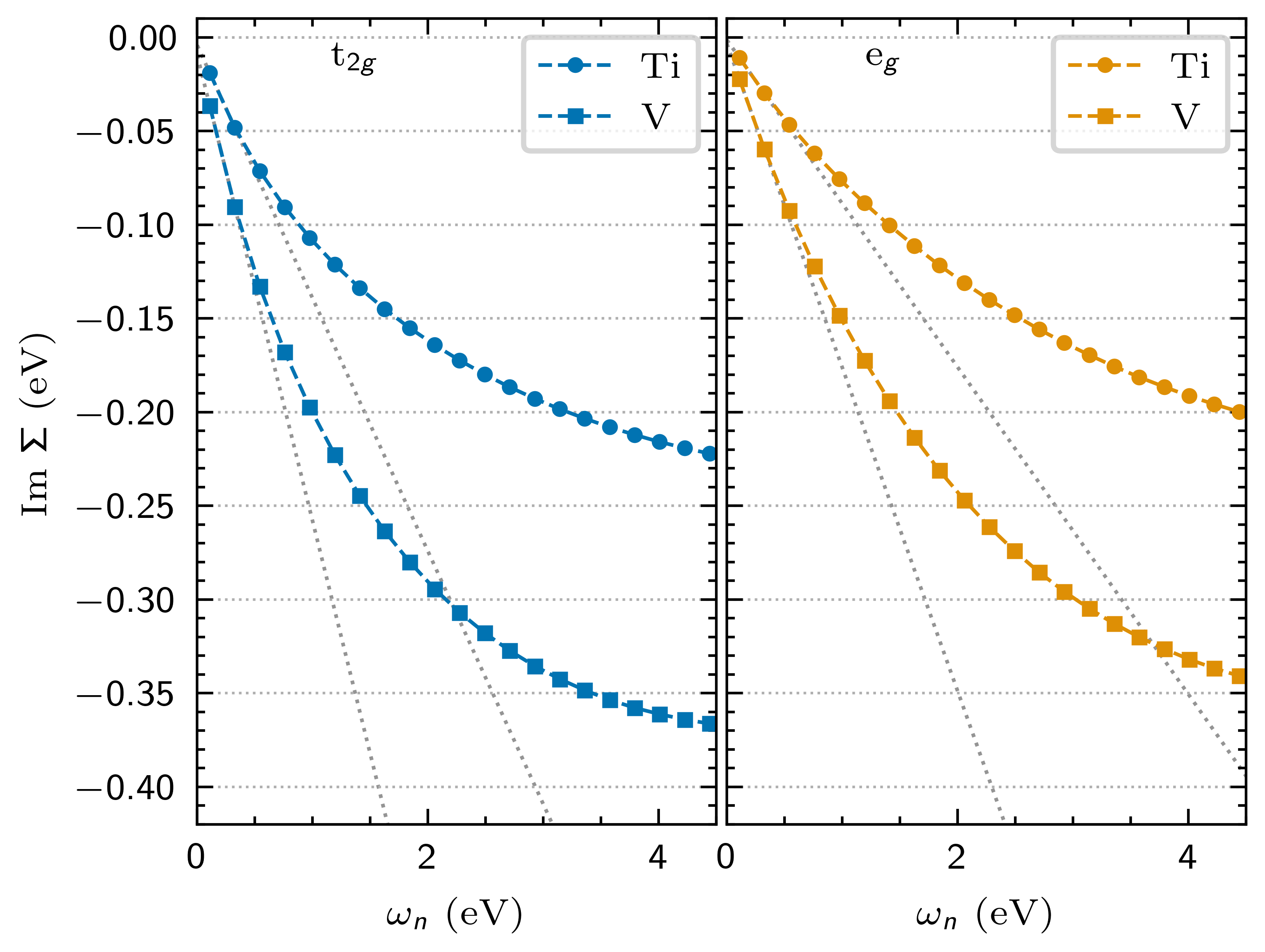}
\caption{Imaginary part of the self-energies as a function of Matsubara frequencies of the t$_{2g}$ and e$_g$ orbitals of Ti$_{0.35}$V$_{0.65}$ 
for $U=\SI{2}{\electronvolt}$, $J=\SI{0.6}{\electronvolt}$ at $\SI{400}{\kelvin}$. 
}\label{fig:Sig_V-Ti}
\end{figure}

In  Fig.~\ref{fig:Sig_V-Ti} we present the imaginary part of the self-energies of V and Ti in the alloy Ti$_{x}$V$_{1-x}$ with $x=0.35$ in the low-energy Matsubara region.
A linear behavior can be seen, which demonstrates that the Fermi liquid character of d-electrons is preserved also in the presence of site-disorder. 
We estimated the quasi-particle weights for the different concentrations for both alloy components. 

\begin{figure}[ht]
\includegraphics[width=0.95\linewidth,clip=true]{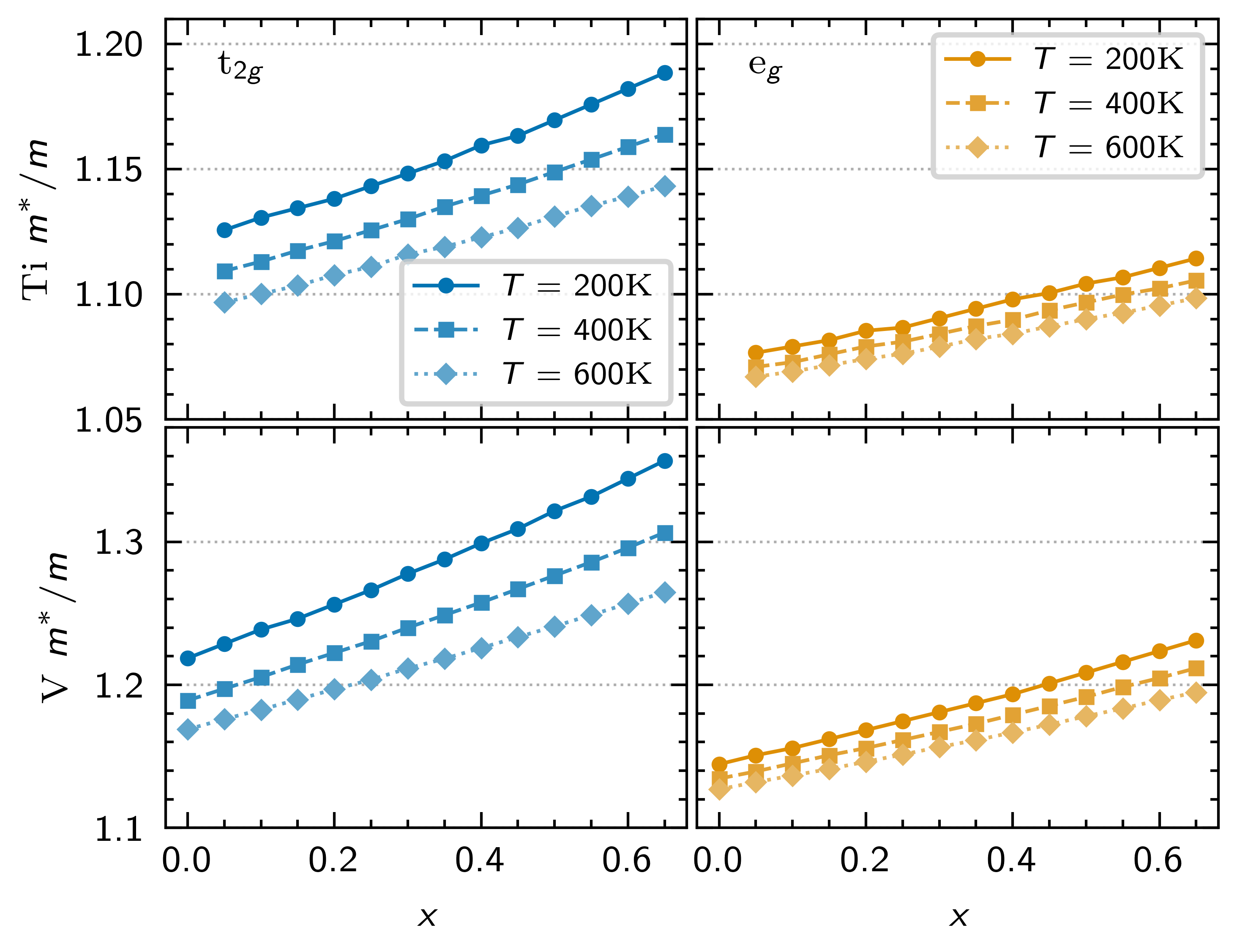}
\caption{Concentration dependence of the effective mass renormalization of the t$_{2g}$ and e$_g$ orbitals for the alloy components Ti and V of Ti$_x$V$_{1-x}$ with $U=\SI{2}{\electronvolt}$ and $J=\SI{0.6}{\electronvolt}$ at different temperatures. 
}\label{fig:meff_V-Ti}
\end{figure}

A quantitative analysis of concentration dependence of the effective masses at temperatures 200, 400 and 600 K, respectively, is presented in Fig.~\ref{fig:meff_V-Ti}. 
The values of the effective mass renormalization $Z^{-1}=m^*/m$ for e$_g$-orbitals of both V and Ti show a very weak concentration dependence and are distributed around a value $Z_{\mathrm{Ti}-e_g}^{-1} \approx 1.07$ and $Z_{\mathrm{V}-e_g}^{-1} \approx 1.16$.
By contrast, the effective mass of the t$_{2g}$-electrons increases slowly upon doping with Ti. 
As the temperature is raised a reduction in the magnitude of the effective masses is also visible.
Note that a reduction of $m^*/m$ is seen also in pristine V, however, the effect is less significant.

\begin{figure}[ht]
\includegraphics[width=0.95\linewidth,clip=true]{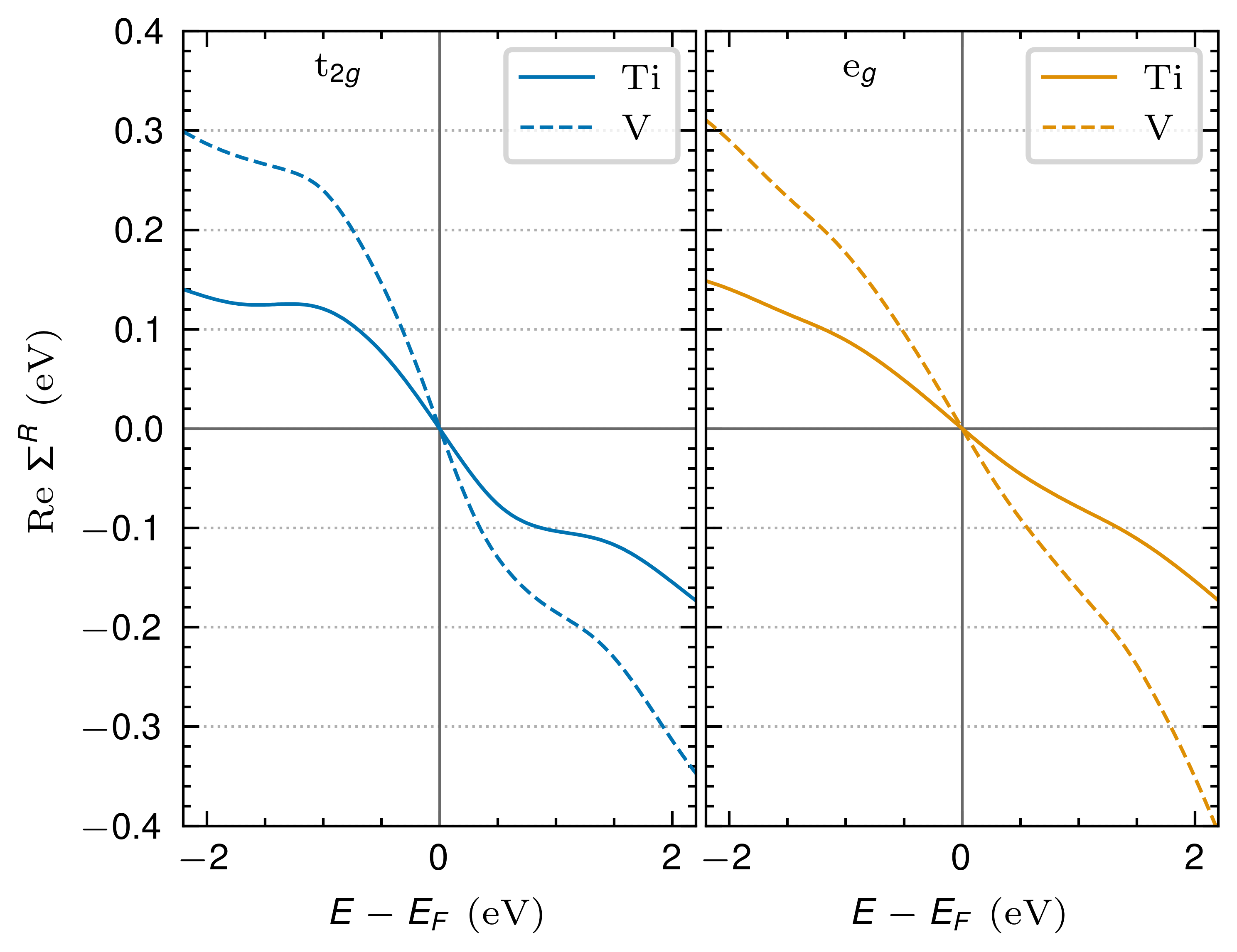}
\caption{Real part of the retarded self-energies as a function of energy of the t$_{2g}$ and e$_g$ orbitals of Ti$_{0.35}$V$_{0.65}$ 
for $U=\SI{2}{\electronvolt}$, $J=\SI{0.6}{\electronvolt}$ at $\SI{400}{\kelvin}$. 
}\label{fig:Sig_z_V-Ti}
\end{figure}

A characteristic phenomenon of disordered and strongly correlated electron systems is the decoherence of quasiparticles. 
In the presence of disorder a resistivity change is observed above some decoherence temperature which is connected with the quasiparticle weight renormalization~\cite{zi.ab.90}. Calculations performed within the framework of DMFT supports this model~\cite{do.ko.93,do.ko.94}. 
Thus, we interpret our LDA+DMFT results for Ti$_x$V$_{1-x}$ alloys using the same picture:
in the presence of a low concentration of Ti impurities at V sites dynamical correlations, through the real part of the self-energy, renormalize the on-site energies (the random on-site potentials): $\epsilon_{t_{2g}/e_g} + \Re \Sigma_{t_{2g}/e_g}^R(E)$. 
For a given energy $E$ in the vicinity of $E_F$ the screening effect is more important for V-d orbitals in comparison with those of Ti-d orbitals. 
In a similar analysis for the orbital resolved self-energies one observes that in the occupied part of the spectra ($E \le E_F$) $\Re \Sigma_{t_{2g}}^R(E) > \Re \Sigma_{e_g}^R(E)$, while in the unoccupied part the spectra the opposite is true: $\Re \Sigma_{t_{2g}}^R(E) < \Re \Sigma_{e_g}^R(E)$. 
Consequently the screening effect is more/less efficient for the t$_{2g}$ orbitals below/above the Fermi level. 
We did not notices significant change in the self-energies Our computations show that as temperature is increasing the screening of the random potential does not change significantly.  
Thus, with increasing the temperatures the t$_{2g}$ quasiparticles scattering become less coherent and the effective masses slightly increase in comparison with the e$_g$ electron quasiparticles.

\subsection{Enhanced critical temperatures in the pure $\beta$- and $\alpha$-phases}

To apply the McMillan formula Eq.~\eqref{eq:mcmilan_tc} the parameters such as the atomic masses and Debye temperatures were taken from Ref.~\cite{cl.fr.60}.
For the latter the concentration dependence have been estimated as the algebraic average of the corresponding Debye temperatures by the relation $\theta_D(x)=x \cdot \theta^{Ti}_D+ (1-x) \cdot \theta^{V}_D$. 
The dependence of the lattice parameter with concentration follows Vegard's law.

\begin{figure}[ht]
\includegraphics[width=0.95\linewidth,clip=true]{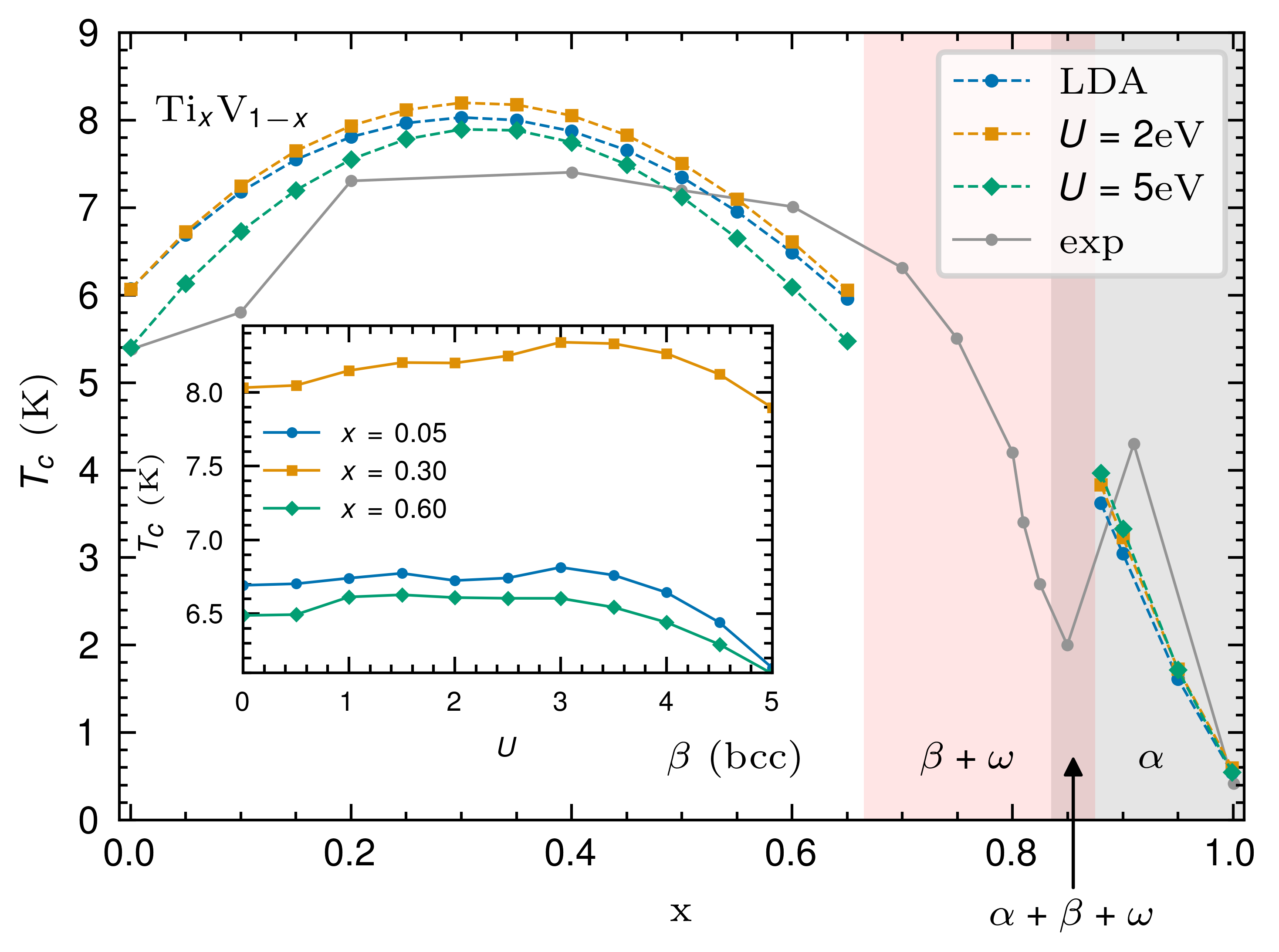}
\caption{Concentration dependence of  $T_c$ in the $\beta$ (bcc) and $\alpha$ (hcp) phases of Ti$_x$V$_{1-x}$. The maximal value for $T_c$ can be observed at $x \approx \num{0.33}$ and is independent of $U$.
The inset shows the interaction strength dependence of $T_c$.
}\label{fig:TC_V-Ti}
\end{figure}

Figure~\ref{fig:TC_V-Ti} shows the concentration dependence of the critical temperature for the Ti$_x$V$_{1-x}$ alloys.
The critical temperature reaches a maximal value (independent of the interaction strength) at a concentration of $x \approx 0.33$.
The inset of Fig.~\ref{fig:TC_V-Ti} shows the critical temperature $T_c$ in dependence of the interaction strength $U$ for the concentration around the maximum of $T_c$ and for two concentrations below and above. Similar to the behavior of pristine vanadium in Fig.~\ref{fig:TcU_V}, the critical temperature rises slightly with increasing $U$ before it drops for strong interactions of about $U > \SI{3}{\electronvolt}$.

LDA+DMFT computations were also performed for the $\alpha$(hexagonal)-phase of Ti$_x$V$_{1-x}$. 
Similar to vanadium, we have optimized the lattice parameter $a$ of the closed-packed hexagonal (hcp) lattice structure, while maintaining a constant ratio of $c/a$.
The McMillan approximation reproduces the increasing trend of the critical temperature $T_C$ for small concentrations $1-x$ of vanadium.
Computations within the mixed $\beta+\omega$- and $\alpha+\beta+\omega$-phase requires a modeling of the crystal structure that cannot be addressed in our current methodology.   

The McMillan formula of $T_c$, Eq.~\eqref{eq:mcmilan_tc}, emerges from the strong-coupling theory with the additional rigid-ion and one-electron approximations. The key physical quantity accessible also through the \textit{ab-initio} computation is the mass enhancement factor $\lambda$ which contains the electron-phonon matrix element averaged over the Fermi surface (see the discussion in Sec.~\ref{sec:Tc}). 

\begin{figure}[ht]
\includegraphics[width=0.95\linewidth,clip=true]{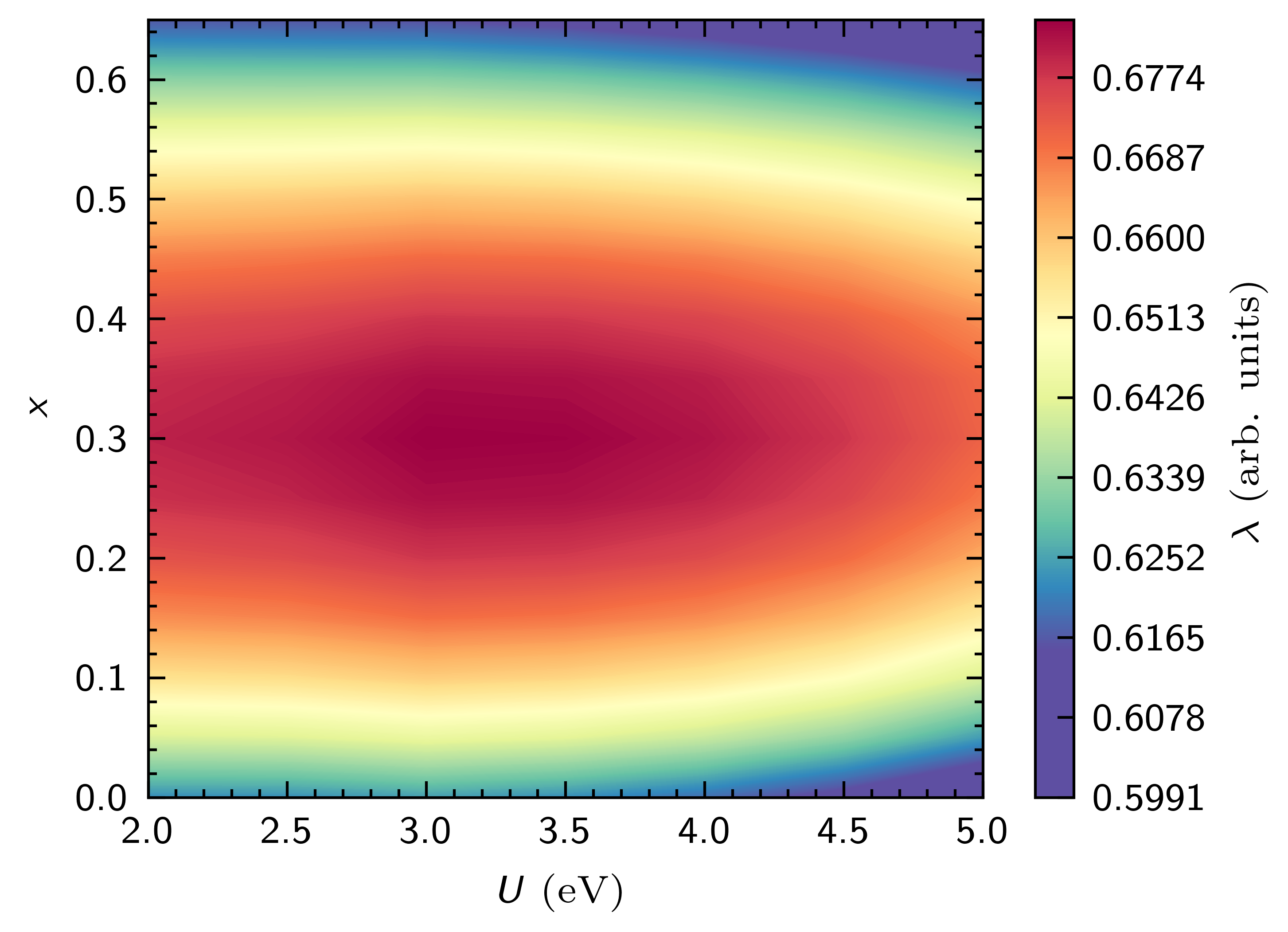}
\caption{Surface plot of $\lambda(U,x)$ for Ti$_x$V$_{1-x}$ alloys computed with LDA+DMFT at $\SI{400}{\kelvin}$.
}\label{fig:TC_V-lambda}
\end{figure}

LDA+DMFT computations show that dynamic electronic correlation in vanadium-titanium alloys leads to narrower DOS around the Fermi level (Fig.~\ref{fig:TiV_DOS}) and an increased  effective mass (see Fig.~\ref{fig:meff_V-Ti}). 
For this reason quasiparticles with energy close to the d-resonance of the electron-ion interaction spends longer times in the vicinity of the scatterer (the so-called Wigner delay-time). 
Therefore, it is reasonable to suppose that the scattering of electrons 
is dominated by the screened local potential, namely, the potential change due to the displacement of an ion and the additional real-part of the local self-energy. 
Thus, the local dynamic correlations lead to a modified electron-phonon coupling which may finally lead to an increase in $T_c$.

Fig.~\ref{fig:TC_V-lambda} shows a surface plot of $\lambda(U,x)$ at temperature $T=\SI{400}{\kelvin}$. For a given $U$ value the $\lambda$ increases, reaching a maximum at a concentration about $x \approx 0.33$ then follows a descending trend. The behavior of $\lambda$ provides a coherent picture since it yields a good estimate for the concentration dependence of the critical temperature seen in Fig.~\ref{fig:TC_V-Ti}.  
 
\section{Summary}
\label{sec:summary}
We computed the superconducting critical temperature of pure V and Ti$_x$V$_{1-x}$ alloys within the $\beta$(bcc)- and $\alpha$-phases, on the basis of BCS-Eliashberg theory.
This theory describes superconductivity caused by an effective electron-electron attraction resulting from the electron-phonon interaction that overcomes the repulsive Coulomb interaction. The magnitude of the critical temperature is limited by the fact that the phonon energy scale is much lower than the electronic energy scales 
(which, in fact, validates the theory).
In particular, we estimated the critical temperatures $T_c$ employing McMillan's formula in which the experimental Debye temperature, the electron-phonon coupling constant $\lambda$ (obtained through \textit{ab-initio} computation), and the pseudo-potential parameter $\mu^*$ are used. We did not attempt to use more advanced formulas for $T_c$ (for example, the Allen-Dynes prescription~\cite{al.dy.75}) involving the phonon band structure and the moments of the Eliashberg function as our aim was to model the enhancement of $T_c$ in Ti$_x$V$_{1-x}$ alloys for which \textit{ab-initio} phonon computations would have involved very expensive supercells. 
Instead, we modeled the dynamical electronic correlations in the studied alloys using the LDA+DMFT method.

As shown previously~\cite{we.be.17} a correct description of the Fermi surface properties of V 
can be achieved on the basis of LDA+DMFT methods which indicated that vanadium is a conventional correlated Fermi liquid. Obviously this does not exclude the possibility that in the regime of very high $T$ and large Hubbard $U$ departures from the Fermi liquid behavior may occur~\cite{be.ka.23}. 
A quantitative comparison with computations involving non-local dynamic correlation effects, LDA+FLEX~\cite{sa.re.18} or GW, showed an excellent agreement
with LDA+DMFT results, in particular, the mass enhancements practically coincide.

Theoretical investigations of disorder in materials including superconductors are usually performed considering the existence of uncorrelated disorder. 
In real systems, however, the disorder is almost never completely random. Inhomogeneities may lead to the appearance of short- or long-range spatial correlations, and modeling of such effects are far more complicated and requires approaches beyond the CPA that are not discussed here.
In our work we considered the regular CPA supplemented by the modeling of dynamical electronic correlations in the normal state of the superconducting Ti$_x$V$_{1-x}$ alloys. 
In particular,
we computed the spectral functions (the \textit{local} DOS) of the alloy component self-energies and the mass enhancements. 

Our critical temperature estimates for the Ti$_x$V$_{1-x}$ alloys with concentrations $x$ in the range of single $\beta$- or $\alpha$-phases are in very good agreement with the experimental measurements. In the rich V region (small $x$) a maximum in the concentration dependence is obtained for $x\approx 0.33$. Although CPA results also indicate such a $T_c$ increase, we show that dynamical electronic correlations modeled with DMFT and a Hubbard $U$ parameter are expected to improve the accuracy of the description.

Early experimental studies interpreted the anomalous behavior of resistivity measurements of certain V-Ti alloys invoking weak localization effects~\cite{sa.mu.90}. 
The recently implemented TMT into the DFT framework allows to search for existence of precursors of localization~\cite{os.zh.20} in alloys with substitutional disorder. 
Within the DFT-TMT framework the \textit{typical} DOS is obtained as a geometrical average of local DOS of alloy components. A vanishing  value of \textit{typical} DOS is considered to signal the presence of weak localization effects~\cite{te.zh.17,os.zh.20,os.ch.21}.  
We have obtained a non-zero \textit{typical} DOS thus, no signature of localization was found for concentrations in which alloys are formed with single phases ($\alpha$ and $\beta$).
Our study does not cover the concentration range $0.65 \le x \le 0.9$ in which a chemically disordered composition with mixed crystal structure phases, $\beta+\omega$ and $\alpha+\beta+\omega$, exists as discussed in the beginning of Sec.~\ref{sec:V-Ti}. 
In this concentration range a direct LDA+DMFT computation is a major challenge because of limitations in computational time and simulation size. We expect that
more advanced multi-scale techniques and possible artificial intelligence methodologies such as machine learning may contribute significantly to data-driven estimations of $T_c$ for disordered superconducting alloys.

\section*{Acknowledgements}
We eminently benefited from discussions with Dieter Vollhardt whose generous
advice is gratefully acknowledged. 
Financial support by the Deutsche Forschungsgemeinschaft (DFG, German Research 
Foundation) through TRR 360, project no. 492547816, is gratefully appreciated.
LC acknowledges the hospitality of Clarendon Laboratory, Department of Physics, University of Oxford, UK, during the final preparation of the manuscript.
LV acknowledges the Swedish Research Council, the Swedish Foundation for Strategic Research, the Carl Trygger Foundation, and the Swedish Foundation for International Cooperation in Research and Higher Education.

\bibliography{sources}

\begin{thebibliography}{78}%
\makeatletter
\providecommand \@ifxundefined [1]{%
 \@ifx{#1\undefined}
}%
\providecommand \@ifnum [1]{%
 \ifnum #1\expandafter \@firstoftwo
 \else \expandafter \@secondoftwo
 \fi
}%
\providecommand \@ifx [1]{%
 \ifx #1\expandafter \@firstoftwo
 \else \expandafter \@secondoftwo
 \fi
}%
\providecommand \natexlab [1]{#1}%
\providecommand \enquote  [1]{``#1''}%
\providecommand \bibnamefont  [1]{#1}%
\providecommand \bibfnamefont [1]{#1}%
\providecommand \citenamefont [1]{#1}%
\providecommand \href@noop [0]{\@secondoftwo}%
\providecommand \href [0]{\begingroup \@sanitize@url \@href}%
\providecommand \@href[1]{\@@startlink{#1}\@@href}%
\providecommand \@@href[1]{\endgroup#1\@@endlink}%
\providecommand \@sanitize@url [0]{\catcode `\\12\catcode `\$12\catcode
  `\&12\catcode `\#12\catcode `\^12\catcode `\_12\catcode `\%12\relax}%
\providecommand \@@startlink[1]{}%
\providecommand \@@endlink[0]{}%
\providecommand \url  [0]{\begingroup\@sanitize@url \@url }%
\providecommand \@url [1]{\endgroup\@href {#1}{\urlprefix }}%
\providecommand \urlprefix  [0]{URL }%
\providecommand \Eprint [0]{\href }%
\providecommand \doibase [0]{https://doi.org/}%
\providecommand \selectlanguage [0]{\@gobble}%
\providecommand \bibinfo  [0]{\@secondoftwo}%
\providecommand \bibfield  [0]{\@secondoftwo}%
\providecommand \translation [1]{[#1]}%
\providecommand \BibitemOpen [0]{}%
\providecommand \bibitemStop [0]{}%
\providecommand \bibitemNoStop [0]{.\EOS\space}%
\providecommand \EOS [0]{\spacefactor3000\relax}%
\providecommand \BibitemShut  [1]{\csname bibitem#1\endcsname}%
\let\auto@bib@innerbib\@empty
\bibitem [{\citenamefont {Bardeen}\ \emph
  {et~al.}(1957{\natexlab{a}})\citenamefont {Bardeen}, \citenamefont {Cooper},\
  and\ \citenamefont {Schrieffer}}]{ba.co.57}%
  \BibitemOpen
  \bibfield  {author} {\bibinfo {author} {\bibfnamefont {J.}~\bibnamefont
  {Bardeen}}, \bibinfo {author} {\bibfnamefont {L.~N.}\ \bibnamefont
  {Cooper}},\ and\ \bibinfo {author} {\bibfnamefont {J.~R.}\ \bibnamefont
  {Schrieffer}},\ }\bibfield  {title} {\bibinfo {title} {Microscopic theory of
  superconductivity},\ }\href {https://doi.org/10.1103/PhysRev.106.162}
  {\bibfield  {journal} {\bibinfo  {journal} {Phys. Rev.}\ }\textbf {\bibinfo
  {volume} {106}},\ \bibinfo {pages} {162} (\bibinfo {year}
  {1957}{\natexlab{a}})}\BibitemShut {NoStop}%
\bibitem [{\citenamefont {Bardeen}\ \emph
  {et~al.}(1957{\natexlab{b}})\citenamefont {Bardeen}, \citenamefont {Cooper},\
  and\ \citenamefont {Schrieffer}}]{ba.co.57b}%
  \BibitemOpen
  \bibfield  {author} {\bibinfo {author} {\bibfnamefont {J.}~\bibnamefont
  {Bardeen}}, \bibinfo {author} {\bibfnamefont {L.~N.}\ \bibnamefont
  {Cooper}},\ and\ \bibinfo {author} {\bibfnamefont {J.~R.}\ \bibnamefont
  {Schrieffer}},\ }\bibfield  {title} {\bibinfo {title} {Theory of
  superconductivity},\ }\href {https://doi.org/10.1103/PhysRev.108.1175}
  {\bibfield  {journal} {\bibinfo  {journal} {Phys. Rev.}\ }\textbf {\bibinfo
  {volume} {108}},\ \bibinfo {pages} {1175} (\bibinfo {year}
  {1957}{\natexlab{b}})}\BibitemShut {NoStop}%
\bibitem [{\citenamefont {Morel}\ and\ \citenamefont
  {Anderson}(1962)}]{mo.an.62}%
  \BibitemOpen
  \bibfield  {author} {\bibinfo {author} {\bibfnamefont {P.}~\bibnamefont
  {Morel}}\ and\ \bibinfo {author} {\bibfnamefont {P.~W.}\ \bibnamefont
  {Anderson}},\ }\bibfield  {title} {\bibinfo {title} {Calculation of the
  superconducting state parameters with retarded electron-phonon interaction},\
  }\href {https://doi.org/10.1103/PhysRev.125.1263} {\bibfield  {journal}
  {\bibinfo  {journal} {Phys. Rev.}\ }\textbf {\bibinfo {volume} {125}},\
  \bibinfo {pages} {1263} (\bibinfo {year} {1962})}\BibitemShut {NoStop}%
\bibitem [{\citenamefont {Eliashberg}(1960)}]{elia.60}%
  \BibitemOpen
  \bibfield  {author} {\bibinfo {author} {\bibfnamefont {G.}~\bibnamefont
  {Eliashberg}},\ }\bibfield  {title} {\bibinfo {title} {Interactions between
  electrons and lattice vibrations in a superconductor},\ }\href@noop {}
  {\bibfield  {journal} {\bibinfo  {journal} {Sov. Phys. JETP}\ }\textbf
  {\bibinfo {volume} {11}},\ \bibinfo {pages} {696} (\bibinfo {year}
  {1960})}\BibitemShut {NoStop}%
\bibitem [{\citenamefont {McMillan}(1968)}]{mcmi.68}%
  \BibitemOpen
  \bibfield  {author} {\bibinfo {author} {\bibfnamefont {W.~L.}\ \bibnamefont
  {McMillan}},\ }\bibfield  {title} {\bibinfo {title} {Transition temperature
  of strong-coupled superconductors},\ }\href
  {https://doi.org/10.1103/PhysRev.167.331} {\bibfield  {journal} {\bibinfo
  {journal} {Phys. Rev.}\ }\textbf {\bibinfo {volume} {167}},\ \bibinfo {pages}
  {331} (\bibinfo {year} {1968})}\BibitemShut {NoStop}%
\bibitem [{\citenamefont {Abrikosov}\ and\ \citenamefont
  {Gor’Kov}(1959)}]{ab.go.59}%
  \BibitemOpen
  \bibfield  {author} {\bibinfo {author} {\bibfnamefont {A.}~\bibnamefont
  {Abrikosov}}\ and\ \bibinfo {author} {\bibfnamefont {L.}~\bibnamefont
  {Gor’Kov}},\ }\bibfield  {title} {\bibinfo {title} {Superconducting alloys
  at finite temperatures},\ }\href@noop {} {\bibfield  {journal} {\bibinfo
  {journal} {Sov. Phys. JETP}\ }\textbf {\bibinfo {volume} {9}},\ \bibinfo
  {pages} {220} (\bibinfo {year} {1959})}\BibitemShut {NoStop}%
\bibitem [{\citenamefont {Anderson}(1959)}]{ande.59}%
  \BibitemOpen
  \bibfield  {author} {\bibinfo {author} {\bibfnamefont {P.~W.}\ \bibnamefont
  {Anderson}},\ }\bibfield  {title} {\bibinfo {title} {Theory of dirty
  superconductors},\ }\href@noop {} {\bibfield  {journal} {\bibinfo  {journal}
  {Journal of Physics and Chemistry of Solids}\ }\textbf {\bibinfo {volume}
  {11}},\ \bibinfo {pages} {26} (\bibinfo {year} {1959})}\BibitemShut {NoStop}%
\bibitem [{\citenamefont {Shahar}\ and\ \citenamefont
  {Ovadyahu}(1992)}]{sh.ov.92}%
  \BibitemOpen
  \bibfield  {author} {\bibinfo {author} {\bibfnamefont {D.}~\bibnamefont
  {Shahar}}\ and\ \bibinfo {author} {\bibfnamefont {Z.}~\bibnamefont
  {Ovadyahu}},\ }\bibfield  {title} {\bibinfo {title} {Superconductivity near
  the mobility edge},\ }\href {https://doi.org/10.1103/PhysRevB.46.10917}
  {\bibfield  {journal} {\bibinfo  {journal} {Phys. Rev. B}\ }\textbf {\bibinfo
  {volume} {46}},\ \bibinfo {pages} {10917} (\bibinfo {year}
  {1992})}\BibitemShut {NoStop}%
\bibitem [{\citenamefont {Sadovskii}(1997)}]{sado.97}%
  \BibitemOpen
  \bibfield  {author} {\bibinfo {author} {\bibfnamefont {M.~V.}\ \bibnamefont
  {Sadovskii}},\ }\bibfield  {title} {\bibinfo {title} {Superconductivity and
  localization},\ }\href@noop {} {\bibfield  {journal} {\bibinfo  {journal}
  {Physics Reports}\ }\textbf {\bibinfo {volume} {282}},\ \bibinfo {pages}
  {225} (\bibinfo {year} {1997})}\BibitemShut {NoStop}%
\bibitem [{\citenamefont {Ghosal}\ \emph {et~al.}(1998)\citenamefont {Ghosal},
  \citenamefont {Randeria},\ and\ \citenamefont {Trivedi}}]{gh.ra.98}%
  \BibitemOpen
  \bibfield  {author} {\bibinfo {author} {\bibfnamefont {A.}~\bibnamefont
  {Ghosal}}, \bibinfo {author} {\bibfnamefont {M.}~\bibnamefont {Randeria}},\
  and\ \bibinfo {author} {\bibfnamefont {N.}~\bibnamefont {Trivedi}},\
  }\bibfield  {title} {\bibinfo {title} {Role of spatial amplitude fluctuations
  in highly disordered $\mathit{s}$-wave superconductors},\ }\href
  {https://doi.org/10.1103/PhysRevLett.81.3940} {\bibfield  {journal} {\bibinfo
   {journal} {Phys. Rev. Lett.}\ }\textbf {\bibinfo {volume} {81}},\ \bibinfo
  {pages} {3940} (\bibinfo {year} {1998})}\BibitemShut {NoStop}%
\bibitem [{\citenamefont {Ghosal}\ \emph {et~al.}(2001)\citenamefont {Ghosal},
  \citenamefont {Randeria},\ and\ \citenamefont {Trivedi}}]{gh.ra.01}%
  \BibitemOpen
  \bibfield  {author} {\bibinfo {author} {\bibfnamefont {A.}~\bibnamefont
  {Ghosal}}, \bibinfo {author} {\bibfnamefont {M.}~\bibnamefont {Randeria}},\
  and\ \bibinfo {author} {\bibfnamefont {N.}~\bibnamefont {Trivedi}},\
  }\bibfield  {title} {\bibinfo {title} {Inhomogeneous pairing in highly
  disordered s-wave superconductors},\ }\href
  {https://doi.org/10.1103/PhysRevB.65.014501} {\bibfield  {journal} {\bibinfo
  {journal} {Phys. Rev. B}\ }\textbf {\bibinfo {volume} {65}},\ \bibinfo
  {pages} {014501} (\bibinfo {year} {2001})}\BibitemShut {NoStop}%
\bibitem [{\citenamefont {Steiner}\ \emph {et~al.}(2008)\citenamefont
  {Steiner}, \citenamefont {Breznay},\ and\ \citenamefont
  {Kapitulnik}}]{st.br.08}%
  \BibitemOpen
  \bibfield  {author} {\bibinfo {author} {\bibfnamefont {M.~A.}\ \bibnamefont
  {Steiner}}, \bibinfo {author} {\bibfnamefont {N.~P.}\ \bibnamefont
  {Breznay}},\ and\ \bibinfo {author} {\bibfnamefont {A.}~\bibnamefont
  {Kapitulnik}},\ }\bibfield  {title} {\bibinfo {title} {Approach to a
  superconductor-to-bose-insulator transition in disordered films},\ }\href
  {https://doi.org/10.1103/PhysRevB.77.212501} {\bibfield  {journal} {\bibinfo
  {journal} {Phys. Rev. B}\ }\textbf {\bibinfo {volume} {77}},\ \bibinfo
  {pages} {212501} (\bibinfo {year} {2008})}\BibitemShut {NoStop}%
\bibitem [{\citenamefont {Sac\'ep\'e}\ \emph {et~al.}(2008)\citenamefont
  {Sac\'ep\'e}, \citenamefont {Chapelier}, \citenamefont {Baturina},
  \citenamefont {Vinokur}, \citenamefont {Baklanov},\ and\ \citenamefont
  {Sanquer}}]{sa.ch.08}%
  \BibitemOpen
  \bibfield  {author} {\bibinfo {author} {\bibfnamefont {B.}~\bibnamefont
  {Sac\'ep\'e}}, \bibinfo {author} {\bibfnamefont {C.}~\bibnamefont
  {Chapelier}}, \bibinfo {author} {\bibfnamefont {T.~I.}\ \bibnamefont
  {Baturina}}, \bibinfo {author} {\bibfnamefont {V.~M.}\ \bibnamefont
  {Vinokur}}, \bibinfo {author} {\bibfnamefont {M.~R.}\ \bibnamefont
  {Baklanov}},\ and\ \bibinfo {author} {\bibfnamefont {M.}~\bibnamefont
  {Sanquer}},\ }\bibfield  {title} {\bibinfo {title} {Disorder-induced
  inhomogeneities of the superconducting state close to the
  superconductor-insulator transition},\ }\href
  {https://doi.org/10.1103/PhysRevLett.101.157006} {\bibfield  {journal}
  {\bibinfo  {journal} {Phys. Rev. Lett.}\ }\textbf {\bibinfo {volume} {101}},\
  \bibinfo {pages} {157006} (\bibinfo {year} {2008})}\BibitemShut {NoStop}%
\bibitem [{\citenamefont {Gantmakher}\ and\ \citenamefont
  {Dolgopolov}(2010)}]{ga.do.10}%
  \BibitemOpen
  \bibfield  {author} {\bibinfo {author} {\bibfnamefont {V.~F.}\ \bibnamefont
  {Gantmakher}}\ and\ \bibinfo {author} {\bibfnamefont {V.~T.}\ \bibnamefont
  {Dolgopolov}},\ }\bibfield  {title} {\bibinfo {title}
  {Superconductor--insulator quantum phase transition},\ }\href@noop {}
  {\bibfield  {journal} {\bibinfo  {journal} {Physics-Uspekhi}\ }\textbf
  {\bibinfo {volume} {53}},\ \bibinfo {pages} {1} (\bibinfo {year}
  {2010})}\BibitemShut {NoStop}%
\bibitem [{\citenamefont {Soven}(1967)}]{sove.67}%
  \BibitemOpen
  \bibfield  {author} {\bibinfo {author} {\bibfnamefont {P.}~\bibnamefont
  {Soven}},\ }\bibfield  {title} {\bibinfo {title} {Coherent-potential model of
  substitutional disordered alloys},\ }\href
  {https://doi.org/10.1103/PhysRev.156.809} {\bibfield  {journal} {\bibinfo
  {journal} {Phys. Rev.}\ }\textbf {\bibinfo {volume} {156}},\ \bibinfo {pages}
  {809} (\bibinfo {year} {1967})}\BibitemShut {NoStop}%
\bibitem [{\citenamefont {Velick\'y}\ \emph {et~al.}(1968)\citenamefont
  {Velick\'y}, \citenamefont {Kirkpatrick},\ and\ \citenamefont
  {Ehrenreich}}]{ve.ki.68}%
  \BibitemOpen
  \bibfield  {author} {\bibinfo {author} {\bibfnamefont {B.}~\bibnamefont
  {Velick\'y}}, \bibinfo {author} {\bibfnamefont {S.}~\bibnamefont
  {Kirkpatrick}},\ and\ \bibinfo {author} {\bibfnamefont {H.}~\bibnamefont
  {Ehrenreich}},\ }\bibfield  {title} {\bibinfo {title} {{Single-Site
  Approximations in the Electronic Theory of Simple Binary Alloys}},\ }\href
  {https://doi.org/10.1103/PhysRev.175.747} {\bibfield  {journal} {\bibinfo
  {journal} {Phys. Rev.}\ }\textbf {\bibinfo {volume} {175}},\ \bibinfo {pages}
  {747} (\bibinfo {year} {1968})}\BibitemShut {NoStop}%
\bibitem [{\citenamefont {Elliott}\ \emph {et~al.}(1974)\citenamefont
  {Elliott}, \citenamefont {Krumhansl},\ and\ \citenamefont
  {Leath}}]{el.kr.74}%
  \BibitemOpen
  \bibfield  {author} {\bibinfo {author} {\bibfnamefont {R.~J.}\ \bibnamefont
  {Elliott}}, \bibinfo {author} {\bibfnamefont {J.~A.}\ \bibnamefont
  {Krumhansl}},\ and\ \bibinfo {author} {\bibfnamefont {P.~L.}\ \bibnamefont
  {Leath}},\ }\bibfield  {title} {\bibinfo {title} {{The theory and properties
  of randomly disordered crystals and related physical systems}},\ }\href
  {https://doi.org/10.1103/RevModPhys.46.465} {\bibfield  {journal} {\bibinfo
  {journal} {Rev. Mod. Phys.}\ }\textbf {\bibinfo {volume} {46}},\ \bibinfo
  {pages} {465} (\bibinfo {year} {1974})}\BibitemShut {NoStop}%
\bibitem [{\citenamefont {Jarrell}\ and\ \citenamefont
  {Krishnamurthy}(2001)}]{ja.kr.01}%
  \BibitemOpen
  \bibfield  {author} {\bibinfo {author} {\bibfnamefont {M.}~\bibnamefont
  {Jarrell}}\ and\ \bibinfo {author} {\bibfnamefont {H.~R.}\ \bibnamefont
  {Krishnamurthy}},\ }\bibfield  {title} {\bibinfo {title} {{Systematic and
  causal corrections to the coherent potential approximation}},\ }\href
  {https://doi.org/10.1103/PhysRevB.63.125102} {\bibfield  {journal} {\bibinfo
  {journal} {Phys. Rev. B}\ }\textbf {\bibinfo {volume} {63}},\ \bibinfo
  {pages} {125102} (\bibinfo {year} {2001})}\BibitemShut {NoStop}%
\bibitem [{\citenamefont {Georges}\ \emph {et~al.}(1996)\citenamefont
  {Georges}, \citenamefont {Kotliar}, \citenamefont {Krauth},\ and\
  \citenamefont {Rozenberg}}]{ge.ko.96}%
  \BibitemOpen
  \bibfield  {author} {\bibinfo {author} {\bibfnamefont {A.}~\bibnamefont
  {Georges}}, \bibinfo {author} {\bibfnamefont {G.}~\bibnamefont {Kotliar}},
  \bibinfo {author} {\bibfnamefont {W.}~\bibnamefont {Krauth}},\ and\ \bibinfo
  {author} {\bibfnamefont {M.~J.}\ \bibnamefont {Rozenberg}},\ }\bibfield
  {title} {\bibinfo {title} {Dynamical mean-field theory of strongly correlated
  fermion systems and the limit of infinite dimensions},\ }\href
  {https://doi.org/10.1103/RevModPhys.68.13} {\bibfield  {journal} {\bibinfo
  {journal} {Rev. Mod. Phys.}\ }\textbf {\bibinfo {volume} {68}},\ \bibinfo
  {pages} {13} (\bibinfo {year} {1996})}\BibitemShut {NoStop}%
\bibitem [{\citenamefont {Kotliar}\ and\ \citenamefont
  {Vollhardt}(2004)}]{ko.vo.04}%
  \BibitemOpen
  \bibfield  {author} {\bibinfo {author} {\bibfnamefont {G.}~\bibnamefont
  {Kotliar}}\ and\ \bibinfo {author} {\bibfnamefont {D.}~\bibnamefont
  {Vollhardt}},\ }\bibfield  {title} {\bibinfo {title} {Strongly correlated
  materials: Insights from dynamical mean-field theory},\ }\href@noop {}
  {\bibfield  {journal} {\bibinfo  {journal} {Physics Today}\ }\textbf
  {\bibinfo {volume} {57}},\ \bibinfo {pages} {53} (\bibinfo {year}
  {2004})}\BibitemShut {NoStop}%
\bibitem [{\citenamefont {Held}(2007)}]{held.07}%
  \BibitemOpen
  \bibfield  {author} {\bibinfo {author} {\bibfnamefont {K.}~\bibnamefont
  {Held}},\ }\bibfield  {title} {\bibinfo {title} {Electronic structure
  calculations using dynamical mean field theory},\ }\href@noop {} {\bibfield
  {journal} {\bibinfo  {journal} {Adv. Phys.}\ }\textbf {\bibinfo {volume}
  {56}},\ \bibinfo {pages} {829} (\bibinfo {year} {2007})}\BibitemShut
  {NoStop}%
\bibitem [{\citenamefont {Dobrosavljevi\'{c}}\ \emph
  {et~al.}(2003)\citenamefont {Dobrosavljevi\'{c}}, \citenamefont {Pastor},\
  and\ \citenamefont {Nikoli\'{c}}}]{do.pa.03}%
  \BibitemOpen
  \bibfield  {author} {\bibinfo {author} {\bibfnamefont {V.}~\bibnamefont
  {Dobrosavljevi\'{c}}}, \bibinfo {author} {\bibfnamefont {A.~A.}\ \bibnamefont
  {Pastor}},\ and\ \bibinfo {author} {\bibfnamefont {B.~K.}\ \bibnamefont
  {Nikoli\'{c}}},\ }\bibfield  {title} {\bibinfo {title} {Typical medium theory
  of anderson localization: A local order parameter approach to strong-disorder
  effects},\ }\href@noop {} {\bibfield  {journal} {\bibinfo  {journal} {EPL}\
  }\textbf {\bibinfo {volume} {62}},\ \bibinfo {pages} {76} (\bibinfo {year}
  {2003})}\BibitemShut {NoStop}%
\bibitem [{\citenamefont {Dobrosavljevi\'{c}}(2010)}]{dobr.10}%
  \BibitemOpen
  \bibfield  {author} {\bibinfo {author} {\bibfnamefont {V.}~\bibnamefont
  {Dobrosavljevi\'{c}}},\ }\bibfield  {title} {\bibinfo {title} {{Typical
  medium theory of Mott-Anderson localization}},\ }\href@noop {} {\bibfield
  {journal} {\bibinfo  {journal} {Int. J. Mod. Phys. B}\ }\textbf {\bibinfo
  {volume} {24}},\ \bibinfo {pages} {1680} (\bibinfo {year}
  {2010})}\BibitemShut {NoStop}%
\bibitem [{\citenamefont {Terletska}\ \emph {et~al.}(2017)\citenamefont
  {Terletska}, \citenamefont {Zhang}, \citenamefont {Chioncel}, \citenamefont
  {Vollhardt},\ and\ \citenamefont {Jarrell}}]{te.zh.17}%
  \BibitemOpen
  \bibfield  {author} {\bibinfo {author} {\bibfnamefont {H.}~\bibnamefont
  {Terletska}}, \bibinfo {author} {\bibfnamefont {Y.}~\bibnamefont {Zhang}},
  \bibinfo {author} {\bibfnamefont {L.}~\bibnamefont {Chioncel}}, \bibinfo
  {author} {\bibfnamefont {D.}~\bibnamefont {Vollhardt}},\ and\ \bibinfo
  {author} {\bibfnamefont {M.}~\bibnamefont {Jarrell}},\ }\bibfield  {title}
  {\bibinfo {title} {Typical-medium multiple-scattering theory for disordered
  systems with anderson localization},\ }\href
  {https://doi.org/10.1103/PhysRevB.95.134204} {\bibfield  {journal} {\bibinfo
  {journal} {Phys. Rev. B}\ }\textbf {\bibinfo {volume} {95}},\ \bibinfo
  {pages} {134204} (\bibinfo {year} {2017})}\BibitemShut {NoStop}%
\bibitem [{\citenamefont {Weh}\ \emph {et~al.}(2021)\citenamefont {Weh},
  \citenamefont {Zhang}, \citenamefont {\"Ostlin}, \citenamefont {Terletska},
  \citenamefont {Bauernfeind}, \citenamefont {Tam}, \citenamefont {Evertz},
  \citenamefont {Byczuk}, \citenamefont {Vollhardt},\ and\ \citenamefont
  {Chioncel}}]{we.zh.21}%
  \BibitemOpen
  \bibfield  {author} {\bibinfo {author} {\bibfnamefont {A.}~\bibnamefont
  {Weh}}, \bibinfo {author} {\bibfnamefont {Y.}~\bibnamefont {Zhang}}, \bibinfo
  {author} {\bibfnamefont {A.}~\bibnamefont {\"Ostlin}}, \bibinfo {author}
  {\bibfnamefont {H.}~\bibnamefont {Terletska}}, \bibinfo {author}
  {\bibfnamefont {D.}~\bibnamefont {Bauernfeind}}, \bibinfo {author}
  {\bibfnamefont {K.-M.}\ \bibnamefont {Tam}}, \bibinfo {author} {\bibfnamefont
  {H.~G.}\ \bibnamefont {Evertz}}, \bibinfo {author} {\bibfnamefont
  {K.}~\bibnamefont {Byczuk}}, \bibinfo {author} {\bibfnamefont
  {D.}~\bibnamefont {Vollhardt}},\ and\ \bibinfo {author} {\bibfnamefont
  {L.}~\bibnamefont {Chioncel}},\ }\bibfield  {title} {\bibinfo {title}
  {Dynamical mean-field theory of the anderson-hubbard model with local and
  nonlocal disorder in tensor formulation},\ }\href
  {https://doi.org/10.1103/PhysRevB.104.045127} {\bibfield  {journal} {\bibinfo
   {journal} {Phys. Rev. B}\ }\textbf {\bibinfo {volume} {104}},\ \bibinfo
  {pages} {045127} (\bibinfo {year} {2021})}\BibitemShut {NoStop}%
\bibitem [{\citenamefont {Min\'ar}\ \emph {et~al.}(2017)\citenamefont
  {Min\'ar}, \citenamefont {Ebert},\ and\ \citenamefont {Chioncel}}]{mi.eb.17}%
  \BibitemOpen
  \bibfield  {author} {\bibinfo {author} {\bibfnamefont {J.}~\bibnamefont
  {Min\'ar}}, \bibinfo {author} {\bibfnamefont {H.}~\bibnamefont {Ebert}},\
  and\ \bibinfo {author} {\bibfnamefont {L.}~\bibnamefont {Chioncel}},\
  }\bibfield  {title} {\bibinfo {title} {A self-consistent, relativistic
  implementation of the lsda+dmft method},\ }\href
  {https://doi.org/10.1140/epjst/e2017-70047-5} {\bibfield  {journal} {\bibinfo
   {journal} {Eur. Phys. J. Special Topics}\ }\textbf {\bibinfo {volume}
  {226}},\ \bibinfo {pages} {2477} (\bibinfo {year} {2017})}\BibitemShut
  {NoStop}%
\bibitem [{\citenamefont {\"Ostlin}\ \emph {et~al.}(2020)\citenamefont
  {\"Ostlin}, \citenamefont {Zhang}, \citenamefont {Terletska}, \citenamefont
  {Beiu\ifmmode~\mbox{\c{s}}\else \c{s}\fi{}eanu}, \citenamefont {Popescu},
  \citenamefont {Byczuk}, \citenamefont {Vitos}, \citenamefont {Jarrell},
  \citenamefont {Vollhardt},\ and\ \citenamefont {Chioncel}}]{os.zh.20}%
  \BibitemOpen
  \bibfield  {author} {\bibinfo {author} {\bibfnamefont {A.}~\bibnamefont
  {\"Ostlin}}, \bibinfo {author} {\bibfnamefont {Y.}~\bibnamefont {Zhang}},
  \bibinfo {author} {\bibfnamefont {H.}~\bibnamefont {Terletska}}, \bibinfo
  {author} {\bibfnamefont {F.}~\bibnamefont {Beiu\ifmmode~\mbox{\c{s}}\else
  \c{s}\fi{}eanu}}, \bibinfo {author} {\bibfnamefont {V.}~\bibnamefont
  {Popescu}}, \bibinfo {author} {\bibfnamefont {K.}~\bibnamefont {Byczuk}},
  \bibinfo {author} {\bibfnamefont {L.}~\bibnamefont {Vitos}}, \bibinfo
  {author} {\bibfnamefont {M.}~\bibnamefont {Jarrell}}, \bibinfo {author}
  {\bibfnamefont {D.}~\bibnamefont {Vollhardt}},\ and\ \bibinfo {author}
  {\bibfnamefont {L.}~\bibnamefont {Chioncel}},\ }\bibfield  {title} {\bibinfo
  {title} {Ab initio typical medium theory of substitutional disorder},\ }\href
  {https://doi.org/10.1103/PhysRevB.101.014210} {\bibfield  {journal} {\bibinfo
   {journal} {Phys. Rev. B}\ }\textbf {\bibinfo {volume} {101}},\ \bibinfo
  {pages} {014210} (\bibinfo {year} {2020})}\BibitemShut {NoStop}%
\bibitem [{\citenamefont {Karabin}\ \emph {et~al.}(2022)\citenamefont
  {Karabin}, \citenamefont {Mondal}, \citenamefont {Östlin}, \citenamefont
  {Ho}, \citenamefont {Dobrosavljevic}, \citenamefont {Tam}, \citenamefont
  {Terletska}, \citenamefont {Chioncel}, \citenamefont {Wang},\ and\
  \citenamefont {Eisenbach}}]{ka.mo.22}%
  \BibitemOpen
  \bibfield  {author} {\bibinfo {author} {\bibfnamefont {M.}~\bibnamefont
  {Karabin}}, \bibinfo {author} {\bibfnamefont {W.~R.}\ \bibnamefont {Mondal}},
  \bibinfo {author} {\bibfnamefont {A.}~\bibnamefont {Östlin}}, \bibinfo
  {author} {\bibfnamefont {W.-G.~D.}\ \bibnamefont {Ho}}, \bibinfo {author}
  {\bibfnamefont {V.}~\bibnamefont {Dobrosavljevic}}, \bibinfo {author}
  {\bibfnamefont {K.-M.}\ \bibnamefont {Tam}}, \bibinfo {author} {\bibfnamefont
  {H.}~\bibnamefont {Terletska}}, \bibinfo {author} {\bibfnamefont
  {L.}~\bibnamefont {Chioncel}}, \bibinfo {author} {\bibfnamefont
  {Y.}~\bibnamefont {Wang}},\ and\ \bibinfo {author} {\bibfnamefont
  {M.}~\bibnamefont {Eisenbach}},\ }\bibfield  {title} {\bibinfo {title} {Ab
  initio approaches to high-entropy alloys: a comparison of cpa, sqs, and
  supercell methods},\ }\href {https://doi.org/10.1007/s10853-022-07186-9}
  {\bibfield  {journal} {\bibinfo  {journal} {Journal of Materials Science}\
  }\textbf {\bibinfo {volume} {57}},\ \bibinfo {pages} {10677 } (\bibinfo
  {year} {2022})}\BibitemShut {NoStop}%
\bibitem [{\citenamefont {Weber}\ \emph {et~al.}(2017)\citenamefont {Weber},
  \citenamefont {Benea}, \citenamefont {Appelt}, \citenamefont {Ceeh},
  \citenamefont {Kreuzpaintner}, \citenamefont {Leitner}, \citenamefont
  {Vollhardt}, \citenamefont {Hugenschmidt},\ and\ \citenamefont
  {Chioncel}}]{we.be.17}%
  \BibitemOpen
  \bibfield  {author} {\bibinfo {author} {\bibfnamefont {J.~A.}\ \bibnamefont
  {Weber}}, \bibinfo {author} {\bibfnamefont {D.}~\bibnamefont {Benea}},
  \bibinfo {author} {\bibfnamefont {W.~H.}\ \bibnamefont {Appelt}}, \bibinfo
  {author} {\bibfnamefont {H.}~\bibnamefont {Ceeh}}, \bibinfo {author}
  {\bibfnamefont {W.}~\bibnamefont {Kreuzpaintner}}, \bibinfo {author}
  {\bibfnamefont {M.}~\bibnamefont {Leitner}}, \bibinfo {author} {\bibfnamefont
  {D.}~\bibnamefont {Vollhardt}}, \bibinfo {author} {\bibfnamefont
  {C.}~\bibnamefont {Hugenschmidt}},\ and\ \bibinfo {author} {\bibfnamefont
  {L.}~\bibnamefont {Chioncel}},\ }\bibfield  {title} {\bibinfo {title}
  {Electronic correlations in vanadium revealed by electron-positron
  annihilation measurements},\ }\href
  {https://doi.org/10.1103/PhysRevB.95.075119} {\bibfield  {journal} {\bibinfo
  {journal} {Phys. Rev. B}\ }\textbf {\bibinfo {volume} {95}},\ \bibinfo
  {pages} {075119} (\bibinfo {year} {2017})}\BibitemShut {NoStop}%
\bibitem [{\citenamefont {Kotliar}\ \emph {et~al.}(2006)\citenamefont
  {Kotliar}, \citenamefont {Savrasov}, \citenamefont {Haule}, \citenamefont
  {Oudovenko}, \citenamefont {Parcollet},\ and\ \citenamefont
  {Marianetti}}]{ko.sa.06}%
  \BibitemOpen
  \bibfield  {author} {\bibinfo {author} {\bibfnamefont {G.}~\bibnamefont
  {Kotliar}}, \bibinfo {author} {\bibfnamefont {S.~Y.}\ \bibnamefont
  {Savrasov}}, \bibinfo {author} {\bibfnamefont {K.}~\bibnamefont {Haule}},
  \bibinfo {author} {\bibfnamefont {V.~S.}\ \bibnamefont {Oudovenko}}, \bibinfo
  {author} {\bibfnamefont {O.}~\bibnamefont {Parcollet}},\ and\ \bibinfo
  {author} {\bibfnamefont {C.~A.}\ \bibnamefont {Marianetti}},\ }\bibfield
  {title} {\bibinfo {title} {Electronic structure calculations with dynamical
  mean-field theory},\ }\href {https://doi.org/10.1103/RevModPhys.78.865}
  {\bibfield  {journal} {\bibinfo  {journal} {Rev. Mod. Phys.}\ }\textbf
  {\bibinfo {volume} {78}},\ \bibinfo {pages} {865} (\bibinfo {year}
  {2006})}\BibitemShut {NoStop}%
\bibitem [{\citenamefont {\"Ostlin}\ \emph {et~al.}(2017)\citenamefont
  {\"Ostlin}, \citenamefont {Vitos},\ and\ \citenamefont
  {Chioncel}}]{os.vi.17}%
  \BibitemOpen
  \bibfield  {author} {\bibinfo {author} {\bibfnamefont {A.}~\bibnamefont
  {\"Ostlin}}, \bibinfo {author} {\bibfnamefont {L.}~\bibnamefont {Vitos}},\
  and\ \bibinfo {author} {\bibfnamefont {L.}~\bibnamefont {Chioncel}},\
  }\bibfield  {title} {\bibinfo {title} {Analytic continuation-free green's
  function approach to correlated electronic structure calculations},\ }\href
  {https://doi.org/10.1103/PhysRevB.96.125156} {\bibfield  {journal} {\bibinfo
  {journal} {Phys. Rev. B}\ }\textbf {\bibinfo {volume} {96}},\ \bibinfo
  {pages} {125156} (\bibinfo {year} {2017})}\BibitemShut {NoStop}%
\bibitem [{\citenamefont {\"Ostlin}\ \emph {et~al.}(2018)\citenamefont
  {\"Ostlin}, \citenamefont {Vitos},\ and\ \citenamefont
  {Chioncel}}]{os.vi.18}%
  \BibitemOpen
  \bibfield  {author} {\bibinfo {author} {\bibfnamefont {A.}~\bibnamefont
  {\"Ostlin}}, \bibinfo {author} {\bibfnamefont {L.}~\bibnamefont {Vitos}},\
  and\ \bibinfo {author} {\bibfnamefont {L.}~\bibnamefont {Chioncel}},\
  }\bibfield  {title} {\bibinfo {title} {Correlated electronic structure with
  uncorrelated disorder},\ }\href {https://doi.org/10.1103/PhysRevB.98.235135}
  {\bibfield  {journal} {\bibinfo  {journal} {Phys. Rev. B}\ }\textbf {\bibinfo
  {volume} {98}},\ \bibinfo {pages} {235135} (\bibinfo {year}
  {2018})}\BibitemShut {NoStop}%
\bibitem [{\citenamefont {Hopfield}(1969)}]{hopf.69}%
  \BibitemOpen
  \bibfield  {author} {\bibinfo {author} {\bibfnamefont {J.~J.}\ \bibnamefont
  {Hopfield}},\ }\bibfield  {title} {\bibinfo {title} {Angular momentum and
  transition-metal superconductivity},\ }\href
  {https://doi.org/10.1103/PhysRev.186.443} {\bibfield  {journal} {\bibinfo
  {journal} {Phys. Rev.}\ }\textbf {\bibinfo {volume} {186}},\ \bibinfo {pages}
  {443} (\bibinfo {year} {1969})}\BibitemShut {NoStop}%
\bibitem [{\citenamefont {Gaspari}\ and\ \citenamefont
  {Gyorffy}(1972)}]{ga.gy.72}%
  \BibitemOpen
  \bibfield  {author} {\bibinfo {author} {\bibfnamefont {G.~D.}\ \bibnamefont
  {Gaspari}}\ and\ \bibinfo {author} {\bibfnamefont {B.~L.}\ \bibnamefont
  {Gyorffy}},\ }\bibfield  {title} {\bibinfo {title} {Electron-phonon
  interactions, $d$ resonances, and superconductivity in transition metals},\
  }\href {https://doi.org/10.1103/PhysRevLett.28.801} {\bibfield  {journal}
  {\bibinfo  {journal} {Phys. Rev. Lett.}\ }\textbf {\bibinfo {volume} {28}},\
  \bibinfo {pages} {801} (\bibinfo {year} {1972})}\BibitemShut {NoStop}%
\bibitem [{\citenamefont {Aryasetiawan}\ \emph {et~al.}(2004)\citenamefont
  {Aryasetiawan}, \citenamefont {Imada}, \citenamefont {Georges}, \citenamefont
  {Kotliar}, \citenamefont {Biermann},\ and\ \citenamefont
  {Lichtenstein}}]{ar.im.04}%
  \BibitemOpen
  \bibfield  {author} {\bibinfo {author} {\bibfnamefont {F.}~\bibnamefont
  {Aryasetiawan}}, \bibinfo {author} {\bibfnamefont {M.}~\bibnamefont {Imada}},
  \bibinfo {author} {\bibfnamefont {A.}~\bibnamefont {Georges}}, \bibinfo
  {author} {\bibfnamefont {G.}~\bibnamefont {Kotliar}}, \bibinfo {author}
  {\bibfnamefont {S.}~\bibnamefont {Biermann}},\ and\ \bibinfo {author}
  {\bibfnamefont {A.~I.}\ \bibnamefont {Lichtenstein}},\ }\bibfield  {title}
  {\bibinfo {title} {Frequency-dependent local interactions and low-energy
  effective models from electronic structure calculations},\ }\href
  {https://doi.org/10.1103/PhysRevB.70.195104} {\bibfield  {journal} {\bibinfo
  {journal} {Phys. Rev. B}\ }\textbf {\bibinfo {volume} {70}},\ \bibinfo
  {pages} {195104} (\bibinfo {year} {2004})}\BibitemShut {NoStop}%
\bibitem [{\citenamefont {Miyake}\ and\ \citenamefont
  {Aryasetiawan}(2008)}]{mi.ar.08}%
  \BibitemOpen
  \bibfield  {author} {\bibinfo {author} {\bibfnamefont {T.}~\bibnamefont
  {Miyake}}\ and\ \bibinfo {author} {\bibfnamefont {F.}~\bibnamefont
  {Aryasetiawan}},\ }\bibfield  {title} {\bibinfo {title} {Screened coulomb
  interaction in the maximally localized wannier basis},\ }\href
  {https://doi.org/10.1103/PhysRevB.77.085122} {\bibfield  {journal} {\bibinfo
  {journal} {Phys. Rev. B}\ }\textbf {\bibinfo {volume} {77}},\ \bibinfo
  {pages} {085122} (\bibinfo {year} {2008})}\BibitemShut {NoStop}%
\bibitem [{\citenamefont {Andersen}\ \emph {et~al.}(1994)\citenamefont
  {Andersen}, \citenamefont {Jepsen},\ and\ \citenamefont
  {Krier}}]{an.je.94.2}%
  \BibitemOpen
  \bibfield  {author} {\bibinfo {author} {\bibfnamefont {O.~K.}\ \bibnamefont
  {Andersen}}, \bibinfo {author} {\bibfnamefont {O.}~\bibnamefont {Jepsen}},\
  and\ \bibinfo {author} {\bibfnamefont {G.}~\bibnamefont {Krier}},\
  }\href@noop {} {\emph {\bibinfo {title} {Lectures on Methods of Electronic
  Structure Calculation}}}\ (\bibinfo  {publisher} {World Scientific},\
  \bibinfo {address} {Singapore},\ \bibinfo {year} {1994})\ p.~\bibinfo {pages}
  {63}\BibitemShut {NoStop}%
\bibitem [{\citenamefont {Vitos}\ \emph {et~al.}(2000)\citenamefont {Vitos},
  \citenamefont {Skriver}, \citenamefont {Johansson},\ and\ \citenamefont
  {Koll\'ar}}]{vi.sk.00}%
  \BibitemOpen
  \bibfield  {author} {\bibinfo {author} {\bibfnamefont {L.}~\bibnamefont
  {Vitos}}, \bibinfo {author} {\bibfnamefont {H.~L.}\ \bibnamefont {Skriver}},
  \bibinfo {author} {\bibfnamefont {B.}~\bibnamefont {Johansson}},\ and\
  \bibinfo {author} {\bibfnamefont {J.}~\bibnamefont {Koll\'ar}},\ }\bibfield
  {title} {\bibinfo {title} {Application of the exact muffin-tin orbitals
  theory: the spherical cell approximation},\ }\href
  {https://doi.org/https://doi.org/10.1016/S0927-0256(99)00098-1} {\bibfield
  {journal} {\bibinfo  {journal} {Comp. Mat. Sci.}\ }\textbf {\bibinfo {volume}
  {18}},\ \bibinfo {pages} {24} (\bibinfo {year} {2000})}\BibitemShut {NoStop}%
\bibitem [{\citenamefont {Vitos}(2001)}]{vito.01}%
  \BibitemOpen
  \bibfield  {author} {\bibinfo {author} {\bibfnamefont {L.}~\bibnamefont
  {Vitos}},\ }\bibfield  {title} {\bibinfo {title} {Total-energy method based
  on the exact muffin-tin orbitals theory},\ }\href
  {https://doi.org/https://doi.org/10.1103/PhysRevB.64.014107} {\bibfield
  {journal} {\bibinfo  {journal} {Phys. Rev. B}\ }\textbf {\bibinfo {volume}
  {64}},\ \bibinfo {pages} {014107} (\bibinfo {year} {2001})}\BibitemShut
  {NoStop}%
\bibitem [{\citenamefont {Lichtenstein}\ and\ \citenamefont
  {Katsnelson}(1998)}]{li.ka.97}%
  \BibitemOpen
  \bibfield  {author} {\bibinfo {author} {\bibfnamefont {A.~I.}\ \bibnamefont
  {Lichtenstein}}\ and\ \bibinfo {author} {\bibfnamefont {M.~I.}\ \bibnamefont
  {Katsnelson}},\ }\bibfield  {title} {\bibinfo {title} {Ab initio calculations
  of quasiparticle band structure in correlated systems: \uppercase{LDA++}
  approach},\ }\href@noop {} {\bibfield  {journal} {\bibinfo  {journal} {Phys.
  Rev. B}\ }\textbf {\bibinfo {volume} {57}},\ \bibinfo {pages} {6884}
  (\bibinfo {year} {1998})}\BibitemShut {NoStop}%
\bibitem [{\citenamefont {Pourovskii}\ \emph {et~al.}(2006)\citenamefont
  {Pourovskii}, \citenamefont {Katsnelson},\ and\ \citenamefont
  {Lichtenstein}}]{po.ka.06}%
  \BibitemOpen
  \bibfield  {author} {\bibinfo {author} {\bibfnamefont {L.~V.}\ \bibnamefont
  {Pourovskii}}, \bibinfo {author} {\bibfnamefont {M.~I.}\ \bibnamefont
  {Katsnelson}},\ and\ \bibinfo {author} {\bibfnamefont {A.~I.}\ \bibnamefont
  {Lichtenstein}},\ }\bibfield  {title} {\bibinfo {title} {Correlation effects
  in electronic structure of ${\mathrm{pucoga}}_{5}$},\ }\href
  {https://doi.org/10.1103/PhysRevB.73.060506} {\bibfield  {journal} {\bibinfo
  {journal} {Phys. Rev. B}\ }\textbf {\bibinfo {volume} {73}},\ \bibinfo
  {pages} {060506} (\bibinfo {year} {2006})}\BibitemShut {NoStop}%
\bibitem [{\citenamefont {Katsnelson}\ \emph {et~al.}(2008)\citenamefont
  {Katsnelson}, \citenamefont {Irkhin}, \citenamefont {Chioncel}, \citenamefont
  {Lichtenstein},\ and\ \citenamefont {de~Groot}}]{ir.ka.08}%
  \BibitemOpen
  \bibfield  {author} {\bibinfo {author} {\bibfnamefont {M.~I.}\ \bibnamefont
  {Katsnelson}}, \bibinfo {author} {\bibfnamefont {V.~Y.}\ \bibnamefont
  {Irkhin}}, \bibinfo {author} {\bibfnamefont {L.}~\bibnamefont {Chioncel}},
  \bibinfo {author} {\bibfnamefont {A.~I.}\ \bibnamefont {Lichtenstein}},\ and\
  \bibinfo {author} {\bibfnamefont {R.~A.}\ \bibnamefont {de~Groot}},\
  }\bibfield  {title} {\bibinfo {title} {Half-metallic ferromagnets: From band
  structure to many-body effects},\ }\href
  {https://doi.org/10.1103/RevModPhys.80.315} {\bibfield  {journal} {\bibinfo
  {journal} {Rev. Mod. Phys.}\ }\textbf {\bibinfo {volume} {80}},\ \bibinfo
  {pages} {315} (\bibinfo {year} {2008})}\BibitemShut {NoStop}%
\bibitem [{\citenamefont {Chioncel}\ \emph {et~al.}(2003)\citenamefont
  {Chioncel}, \citenamefont {Vitos}, \citenamefont {Abrikosov}, \citenamefont
  {Koll\'ar}, \citenamefont {Katsnelson},\ and\ \citenamefont
  {Lichtenstein}}]{ch.vi.03}%
  \BibitemOpen
  \bibfield  {author} {\bibinfo {author} {\bibfnamefont {L.}~\bibnamefont
  {Chioncel}}, \bibinfo {author} {\bibfnamefont {L.}~\bibnamefont {Vitos}},
  \bibinfo {author} {\bibfnamefont {I.~A.}\ \bibnamefont {Abrikosov}}, \bibinfo
  {author} {\bibfnamefont {J.}~\bibnamefont {Koll\'ar}}, \bibinfo {author}
  {\bibfnamefont {M.~I.}\ \bibnamefont {Katsnelson}},\ and\ \bibinfo {author}
  {\bibfnamefont {A.~I.}\ \bibnamefont {Lichtenstein}},\ }\bibfield  {title}
  {\bibinfo {title} {Ab initio electronic structure calculations of correlated
  systems: An emto-dmft approach},\ }\href
  {https://doi.org/10.1103/PhysRevB.67.235106} {\bibfield  {journal} {\bibinfo
  {journal} {Phys. Rev. B}\ }\textbf {\bibinfo {volume} {67}},\ \bibinfo
  {pages} {235106} (\bibinfo {year} {2003})}\BibitemShut {NoStop}%
\bibitem [{\citenamefont {\"Ostlin}\ and\ \citenamefont
  {Chioncel}(2021)}]{os.ch.21}%
  \BibitemOpen
  \bibfield  {author} {\bibinfo {author} {\bibfnamefont {A.}~\bibnamefont
  {\"Ostlin}}\ and\ \bibinfo {author} {\bibfnamefont {L.}~\bibnamefont
  {Chioncel}},\ }\bibfield  {title} {\bibinfo {title} {Electronic correlations
  and fermi liquid behavior of intermediate-band states in titanium-doped
  silicon},\ }\href {https://doi.org/10.1103/PhysRevB.104.L201201} {\bibfield
  {journal} {\bibinfo  {journal} {Phys. Rev. B}\ }\textbf {\bibinfo {volume}
  {104}},\ \bibinfo {pages} {L201201} (\bibinfo {year} {2021})}\BibitemShut
  {NoStop}%
\bibitem [{\citenamefont {Marques}\ \emph {et~al.}(2005)\citenamefont
  {Marques}, \citenamefont {L\"uders}, \citenamefont {Lathiotakis},
  \citenamefont {Profeta}, \citenamefont {Floris}, \citenamefont {Fast},
  \citenamefont {Continenza}, \citenamefont {Gross},\ and\ \citenamefont
  {Massidda}}]{ma.lu.05}%
  \BibitemOpen
  \bibfield  {author} {\bibinfo {author} {\bibfnamefont {M.~A.~L.}\
  \bibnamefont {Marques}}, \bibinfo {author} {\bibfnamefont {M.}~\bibnamefont
  {L\"uders}}, \bibinfo {author} {\bibfnamefont {N.~N.}\ \bibnamefont
  {Lathiotakis}}, \bibinfo {author} {\bibfnamefont {G.}~\bibnamefont
  {Profeta}}, \bibinfo {author} {\bibfnamefont {A.}~\bibnamefont {Floris}},
  \bibinfo {author} {\bibfnamefont {L.}~\bibnamefont {Fast}}, \bibinfo {author}
  {\bibfnamefont {A.}~\bibnamefont {Continenza}}, \bibinfo {author}
  {\bibfnamefont {E.~K.~U.}\ \bibnamefont {Gross}},\ and\ \bibinfo {author}
  {\bibfnamefont {S.}~\bibnamefont {Massidda}},\ }\bibfield  {title} {\bibinfo
  {title} {Ab initio theory of superconductivity. ii. application to elemental
  metals},\ }\href {https://doi.org/10.1103/PhysRevB.72.024546} {\bibfield
  {journal} {\bibinfo  {journal} {Phys. Rev. B}\ }\textbf {\bibinfo {volume}
  {72}},\ \bibinfo {pages} {024546} (\bibinfo {year} {2005})}\BibitemShut
  {NoStop}%
\bibitem [{\citenamefont {Gonze}\ \emph {et~al.}(2009)\citenamefont {Gonze},
  \citenamefont {Amadon}, \citenamefont {Anglade}, \citenamefont {Beuken},
  \citenamefont {Bottin}, \citenamefont {Boulanger}, \citenamefont {Bruneval},
  \citenamefont {Caliste}, \citenamefont {Caracas}, \citenamefont {Côté},
  \citenamefont {Deutsch}, \citenamefont {Genovese}, \citenamefont {Ghosez},
  \citenamefont {Giantomassi}, \citenamefont {Goedecker}, \citenamefont
  {Hamann}, \citenamefont {Hermet}, \citenamefont {Jollet}, \citenamefont
  {Jomard}, \citenamefont {Leroux}, \citenamefont {Mancini}, \citenamefont
  {Mazevet}, \citenamefont {Oliveira}, \citenamefont {Onida}, \citenamefont
  {Pouillon}, \citenamefont {Rangel}, \citenamefont {Rignanese}, \citenamefont
  {Sangalli}, \citenamefont {Shaltaf}, \citenamefont {Torrent}, \citenamefont
  {Verstraete}, \citenamefont {Zerah},\ and\ \citenamefont
  {Zwanziger}}]{go.am.09}%
  \BibitemOpen
  \bibfield  {author} {\bibinfo {author} {\bibfnamefont {X.}~\bibnamefont
  {Gonze}}, \bibinfo {author} {\bibfnamefont {B.}~\bibnamefont {Amadon}},
  \bibinfo {author} {\bibfnamefont {P.-M.}\ \bibnamefont {Anglade}}, \bibinfo
  {author} {\bibfnamefont {J.-M.}\ \bibnamefont {Beuken}}, \bibinfo {author}
  {\bibfnamefont {F.}~\bibnamefont {Bottin}}, \bibinfo {author} {\bibfnamefont
  {P.}~\bibnamefont {Boulanger}}, \bibinfo {author} {\bibfnamefont
  {F.}~\bibnamefont {Bruneval}}, \bibinfo {author} {\bibfnamefont
  {D.}~\bibnamefont {Caliste}}, \bibinfo {author} {\bibfnamefont
  {R.}~\bibnamefont {Caracas}}, \bibinfo {author} {\bibfnamefont
  {M.}~\bibnamefont {Côté}}, \bibinfo {author} {\bibfnamefont
  {T.}~\bibnamefont {Deutsch}}, \bibinfo {author} {\bibfnamefont
  {L.}~\bibnamefont {Genovese}}, \bibinfo {author} {\bibfnamefont
  {P.}~\bibnamefont {Ghosez}}, \bibinfo {author} {\bibfnamefont
  {M.}~\bibnamefont {Giantomassi}}, \bibinfo {author} {\bibfnamefont
  {S.}~\bibnamefont {Goedecker}}, \bibinfo {author} {\bibfnamefont
  {D.}~\bibnamefont {Hamann}}, \bibinfo {author} {\bibfnamefont
  {P.}~\bibnamefont {Hermet}}, \bibinfo {author} {\bibfnamefont
  {F.}~\bibnamefont {Jollet}}, \bibinfo {author} {\bibfnamefont
  {G.}~\bibnamefont {Jomard}}, \bibinfo {author} {\bibfnamefont
  {S.}~\bibnamefont {Leroux}}, \bibinfo {author} {\bibfnamefont
  {M.}~\bibnamefont {Mancini}}, \bibinfo {author} {\bibfnamefont
  {S.}~\bibnamefont {Mazevet}}, \bibinfo {author} {\bibfnamefont
  {M.}~\bibnamefont {Oliveira}}, \bibinfo {author} {\bibfnamefont
  {G.}~\bibnamefont {Onida}}, \bibinfo {author} {\bibfnamefont
  {Y.}~\bibnamefont {Pouillon}}, \bibinfo {author} {\bibfnamefont
  {T.}~\bibnamefont {Rangel}}, \bibinfo {author} {\bibfnamefont {G.-M.}\
  \bibnamefont {Rignanese}}, \bibinfo {author} {\bibfnamefont {D.}~\bibnamefont
  {Sangalli}}, \bibinfo {author} {\bibfnamefont {R.}~\bibnamefont {Shaltaf}},
  \bibinfo {author} {\bibfnamefont {M.}~\bibnamefont {Torrent}}, \bibinfo
  {author} {\bibfnamefont {M.}~\bibnamefont {Verstraete}}, \bibinfo {author}
  {\bibfnamefont {G.}~\bibnamefont {Zerah}},\ and\ \bibinfo {author}
  {\bibfnamefont {J.}~\bibnamefont {Zwanziger}},\ }\bibfield  {title} {\bibinfo
  {title} {Abinit: First-principles approach to material and nanosystem
  properties},\ }\href
  {https://doi.org/https://doi.org/10.1016/j.cpc.2009.07.007} {\bibfield
  {journal} {\bibinfo  {journal} {Computer Physics Communications}\ }\textbf
  {\bibinfo {volume} {180}},\ \bibinfo {pages} {2582} (\bibinfo {year}
  {2009})},\ \bibinfo {note} {40 YEARS OF CPC: A celebratory issue focused on
  quality software for high performance, grid and novel computing
  architectures}\BibitemShut {NoStop}%
\bibitem [{\citenamefont {Noffsinger}\ \emph {et~al.}(2010)\citenamefont
  {Noffsinger}, \citenamefont {Giustino}, \citenamefont {Malone}, \citenamefont
  {Park}, \citenamefont {Louie},\ and\ \citenamefont {Cohen}}]{no.gi.10}%
  \BibitemOpen
  \bibfield  {author} {\bibinfo {author} {\bibfnamefont {J.}~\bibnamefont
  {Noffsinger}}, \bibinfo {author} {\bibfnamefont {F.}~\bibnamefont
  {Giustino}}, \bibinfo {author} {\bibfnamefont {B.~D.}\ \bibnamefont
  {Malone}}, \bibinfo {author} {\bibfnamefont {C.-H.}\ \bibnamefont {Park}},
  \bibinfo {author} {\bibfnamefont {S.~G.}\ \bibnamefont {Louie}},\ and\
  \bibinfo {author} {\bibfnamefont {M.~L.}\ \bibnamefont {Cohen}},\ }\bibfield
  {title} {\bibinfo {title} {Epw: A program for calculating the
  electron–phonon coupling using maximally localized wannier functions},\
  }\href {https://doi.org/https://doi.org/10.1016/j.cpc.2010.08.027} {\bibfield
   {journal} {\bibinfo  {journal} {Computer Physics Communications}\ }\textbf
  {\bibinfo {volume} {181}},\ \bibinfo {pages} {2140} (\bibinfo {year}
  {2010})}\BibitemShut {NoStop}%
\bibitem [{\citenamefont {Gomersall}\ and\ \citenamefont
  {Gyorffy}(1974)}]{go.gy.74}%
  \BibitemOpen
  \bibfield  {author} {\bibinfo {author} {\bibfnamefont {I.~R.}\ \bibnamefont
  {Gomersall}}\ and\ \bibinfo {author} {\bibfnamefont {B.~L.}\ \bibnamefont
  {Gyorffy}},\ }\bibfield  {title} {\bibinfo {title} {A simple theory of the
  electron-phonon mass enhancement in transition metal compounds},\ }\href
  {https://doi.org/10.1088/0305-4608/4/8/015} {\bibfield  {journal} {\bibinfo
  {journal} {Journal of Physics F: Metal Physics}\ }\textbf {\bibinfo {volume}
  {4}},\ \bibinfo {pages} {1204} (\bibinfo {year} {1974})}\BibitemShut
  {NoStop}%
\bibitem [{\citenamefont {Vonsovsky}\ \emph {et~al.}(2011)\citenamefont
  {Vonsovsky}, \citenamefont {Brandt}, \citenamefont {Izyumov}, \citenamefont
  {Zavarnitsyn},\ and\ \citenamefont {Kurmaev}}]{vo.br.11}%
  \BibitemOpen
  \bibfield  {author} {\bibinfo {author} {\bibfnamefont {S.}~\bibnamefont
  {Vonsovsky}}, \bibinfo {author} {\bibfnamefont {E.}~\bibnamefont {Brandt}},
  \bibinfo {author} {\bibfnamefont {Y.}~\bibnamefont {Izyumov}}, \bibinfo
  {author} {\bibfnamefont {A.}~\bibnamefont {Zavarnitsyn}},\ and\ \bibinfo
  {author} {\bibfnamefont {E.}~\bibnamefont {Kurmaev}},\ }\href
  {https://books.google.de/books?id=ANOAMQEACAAJ} {\emph {\bibinfo {title}
  {Superconductivity of Transition Metals: Their Alloys and Compounds}}},\
  Springer Series in Solid-State Sciences\ (\bibinfo  {publisher} {Springer
  Berlin Heidelberg},\ \bibinfo {year} {2011})\BibitemShut {NoStop}%
\bibitem [{\citenamefont {Westlake}\ and\ \citenamefont
  {Alfred}(1968)}]{we.al.68}%
  \BibitemOpen
  \bibfield  {author} {\bibinfo {author} {\bibfnamefont {D.}~\bibnamefont
  {Westlake}}\ and\ \bibinfo {author} {\bibfnamefont {L.}~\bibnamefont
  {Alfred}},\ }\bibfield  {title} {\bibinfo {title} {Determination of the debye
  characteristic temperature of vanadium from the bloch-grüneisen relation},\
  }\href {https://doi.org/https://doi.org/10.1016/0022-3697(68)90043-7}
  {\bibfield  {journal} {\bibinfo  {journal} {Journal of Physics and Chemistry
  of Solids}\ }\textbf {\bibinfo {volume} {29}},\ \bibinfo {pages} {1931}
  (\bibinfo {year} {1968})}\BibitemShut {NoStop}%
\bibitem [{\citenamefont {Tokii}\ and\ \citenamefont {Wakoh}(2003)}]{to.wa.03}%
  \BibitemOpen
  \bibfield  {author} {\bibinfo {author} {\bibfnamefont {M.}~\bibnamefont
  {Tokii}}\ and\ \bibinfo {author} {\bibfnamefont {S.}~\bibnamefont {Wakoh}},\
  }\bibfield  {title} {\bibinfo {title} {An effective potential of metallic
  vanadium and chromium},\ }\href {https://doi.org/10.1143/JPSJ.72.1476}
  {\bibfield  {journal} {\bibinfo  {journal} {Journal of the Physical Society
  of Japan}\ }\textbf {\bibinfo {volume} {72}},\ \bibinfo {pages} {1476}
  (\bibinfo {year} {2003})},\ \Eprint
  {https://arxiv.org/abs/https://doi.org/10.1143/JPSJ.72.1476}
  {https://doi.org/10.1143/JPSJ.72.1476} \BibitemShut {NoStop}%
\bibitem [{\citenamefont {Belozerov}\ \emph {et~al.}(2023)\citenamefont
  {Belozerov}, \citenamefont {Katanin},\ and\ \citenamefont
  {Anisimov}}]{be.ka.23}%
  \BibitemOpen
  \bibfield  {author} {\bibinfo {author} {\bibfnamefont {A.~S.}\ \bibnamefont
  {Belozerov}}, \bibinfo {author} {\bibfnamefont {A.~A.}\ \bibnamefont
  {Katanin}},\ and\ \bibinfo {author} {\bibfnamefont {V.~I.}\ \bibnamefont
  {Anisimov}},\ }\bibfield  {title} {\bibinfo {title} {Transition from pauli
  paramagnetism to curie-weiss behavior in vanadium},\ }\href
  {https://doi.org/10.1103/PhysRevB.107.035116} {\bibfield  {journal} {\bibinfo
   {journal} {Phys. Rev. B}\ }\textbf {\bibinfo {volume} {107}},\ \bibinfo
  {pages} {035116} (\bibinfo {year} {2023})}\BibitemShut {NoStop}%
\bibitem [{\citenamefont {Sihi}\ and\ \citenamefont {Pandey}(2020)}]{si.pa.20}%
  \BibitemOpen
  \bibfield  {author} {\bibinfo {author} {\bibfnamefont {A.}~\bibnamefont
  {Sihi}}\ and\ \bibinfo {author} {\bibfnamefont {S.~K.}\ \bibnamefont
  {Pandey}},\ }\bibfield  {title} {\bibinfo {title} {A detailed electronic
  structure study of vanadium metal by using different beyond-dft methods},\
  }\bibfield  {journal} {\bibinfo  {journal} {The European Physical Journal B}\
  }\textbf {\bibinfo {volume} {93}},\ \href
  {https://doi.org/10.1140/epjb/e2019-100500-8} {10.1140/epjb/e2019-100500-8}
  (\bibinfo {year} {2020})\BibitemShut {NoStop}%
\bibitem [{\citenamefont {Savrasov}\ \emph {et~al.}(2018)\citenamefont
  {Savrasov}, \citenamefont {Resta},\ and\ \citenamefont {Wan}}]{sa.re.18}%
  \BibitemOpen
  \bibfield  {author} {\bibinfo {author} {\bibfnamefont {S.~Y.}\ \bibnamefont
  {Savrasov}}, \bibinfo {author} {\bibfnamefont {G.}~\bibnamefont {Resta}},\
  and\ \bibinfo {author} {\bibfnamefont {X.}~\bibnamefont {Wan}},\ }\bibfield
  {title} {\bibinfo {title} {Local self-energies for v and pd emergent from a
  nonlocal lda+flex implementation},\ }\href
  {https://doi.org/10.1103/PhysRevB.97.155128} {\bibfield  {journal} {\bibinfo
  {journal} {Phys. Rev. B}\ }\textbf {\bibinfo {volume} {97}},\ \bibinfo
  {pages} {155128} (\bibinfo {year} {2018})}\BibitemShut {NoStop}%
\bibitem [{\citenamefont {Louis}\ and\ \citenamefont
  {Iyakutti}(2003)}]{lo.iy.03}%
  \BibitemOpen
  \bibfield  {author} {\bibinfo {author} {\bibfnamefont {C.~N.}\ \bibnamefont
  {Louis}}\ and\ \bibinfo {author} {\bibfnamefont {K.}~\bibnamefont
  {Iyakutti}},\ }\bibfield  {title} {\bibinfo {title} {Electronic phase
  transition and superconductivity of vanadium under high pressure},\ }\href
  {https://doi.org/10.1103/PhysRevB.67.094509} {\bibfield  {journal} {\bibinfo
  {journal} {Phys. Rev. B}\ }\textbf {\bibinfo {volume} {67}},\ \bibinfo
  {pages} {094509} (\bibinfo {year} {2003})}\BibitemShut {NoStop}%
\bibitem [{\citenamefont {{Matin}}\ \emph {et~al.}(2014)\citenamefont
  {{Matin}}, \citenamefont {{Sharath Chandra}}, \citenamefont {{Pandey}},
  \citenamefont {{Chattopadhyay}},\ and\ \citenamefont {{Roy}}}]{ma.sh.14}%
  \BibitemOpen
  \bibfield  {author} {\bibinfo {author} {\bibfnamefont {M.}~\bibnamefont
  {{Matin}}}, \bibinfo {author} {\bibfnamefont {L.~S.}\ \bibnamefont {{Sharath
  Chandra}}}, \bibinfo {author} {\bibfnamefont {S.~K.}\ \bibnamefont
  {{Pandey}}}, \bibinfo {author} {\bibfnamefont {M.~K.}\ \bibnamefont
  {{Chattopadhyay}}},\ and\ \bibinfo {author} {\bibfnamefont {S.~B.}\
  \bibnamefont {{Roy}}},\ }\bibfield  {title} {\bibinfo {title} {{The influence
  of electron-phonon coupling and spin fluctuations on the superconductivity of
  the Ti-V alloys}},\ }\href {https://doi.org/10.1140/epjb/e2014-50036-2}
  {\bibfield  {journal} {\bibinfo  {journal} {European Physical Journal B}\
  }\textbf {\bibinfo {volume} {87}},\ \bibinfo {eid} {131} (\bibinfo {year}
  {2014})}\BibitemShut {NoStop}%
\bibitem [{\citenamefont {Berk}\ and\ \citenamefont
  {Schrieffer}(1966)}]{be.sc.66}%
  \BibitemOpen
  \bibfield  {author} {\bibinfo {author} {\bibfnamefont {N.~F.}\ \bibnamefont
  {Berk}}\ and\ \bibinfo {author} {\bibfnamefont {J.~R.}\ \bibnamefont
  {Schrieffer}},\ }\bibfield  {title} {\bibinfo {title} {Effect of
  ferromagnetic spin correlations on superconductivity},\ }\href
  {https://doi.org/10.1103/PhysRevLett.17.433} {\bibfield  {journal} {\bibinfo
  {journal} {Phys. Rev. Lett.}\ }\textbf {\bibinfo {volume} {17}},\ \bibinfo
  {pages} {433} (\bibinfo {year} {1966})}\BibitemShut {NoStop}%
\bibitem [{\citenamefont {Rietschel}\ and\ \citenamefont
  {Winter}(1979)}]{ri.wi.79}%
  \BibitemOpen
  \bibfield  {author} {\bibinfo {author} {\bibfnamefont {H.}~\bibnamefont
  {Rietschel}}\ and\ \bibinfo {author} {\bibfnamefont {H.}~\bibnamefont
  {Winter}},\ }\bibfield  {title} {\bibinfo {title} {Role of spin fluctuations
  in the superconductors nb and v},\ }\href
  {https://doi.org/10.1103/PhysRevLett.43.1256} {\bibfield  {journal} {\bibinfo
   {journal} {Phys. Rev. Lett.}\ }\textbf {\bibinfo {volume} {43}},\ \bibinfo
  {pages} {1256} (\bibinfo {year} {1979})}\BibitemShut {NoStop}%
\bibitem [{\citenamefont {Ley}\ \emph {et~al.}(1977)\citenamefont {Ley},
  \citenamefont {Dabbousi}, \citenamefont {Kowalczyk}, \citenamefont
  {McFeely},\ and\ \citenamefont {Shirley}}]{le.da.77}%
  \BibitemOpen
  \bibfield  {author} {\bibinfo {author} {\bibfnamefont {L.}~\bibnamefont
  {Ley}}, \bibinfo {author} {\bibfnamefont {O.~B.}\ \bibnamefont {Dabbousi}},
  \bibinfo {author} {\bibfnamefont {S.~P.}\ \bibnamefont {Kowalczyk}}, \bibinfo
  {author} {\bibfnamefont {F.~R.}\ \bibnamefont {McFeely}},\ and\ \bibinfo
  {author} {\bibfnamefont {D.~A.}\ \bibnamefont {Shirley}},\ }\bibfield
  {title} {\bibinfo {title} {X-ray photoemission spectra of the valence bands
  of the $3d$ transition metals, sc to fe},\ }\href
  {https://doi.org/10.1103/PhysRevB.16.5372} {\bibfield  {journal} {\bibinfo
  {journal} {Phys. Rev. B}\ }\textbf {\bibinfo {volume} {16}},\ \bibinfo
  {pages} {5372} (\bibinfo {year} {1977})}\BibitemShut {NoStop}%
\bibitem [{\citenamefont {Speier}\ \emph {et~al.}(1984)\citenamefont {Speier},
  \citenamefont {Fuggle}, \citenamefont {Zeller}, \citenamefont {Ackermann},
  \citenamefont {Szot}, \citenamefont {Hillebrecht},\ and\ \citenamefont
  {Campagna}}]{sp.fu.84}%
  \BibitemOpen
  \bibfield  {author} {\bibinfo {author} {\bibfnamefont {W.}~\bibnamefont
  {Speier}}, \bibinfo {author} {\bibfnamefont {J.~C.}\ \bibnamefont {Fuggle}},
  \bibinfo {author} {\bibfnamefont {R.}~\bibnamefont {Zeller}}, \bibinfo
  {author} {\bibfnamefont {B.}~\bibnamefont {Ackermann}}, \bibinfo {author}
  {\bibfnamefont {K.}~\bibnamefont {Szot}}, \bibinfo {author} {\bibfnamefont
  {F.~U.}\ \bibnamefont {Hillebrecht}},\ and\ \bibinfo {author} {\bibfnamefont
  {M.}~\bibnamefont {Campagna}},\ }\bibfield  {title} {\bibinfo {title}
  {Bremsstrahlung isochromat spectra and density-of-states calculations for the
  $3d$ and $4d$ transition metals},\ }\href
  {https://doi.org/10.1103/PhysRevB.30.6921} {\bibfield  {journal} {\bibinfo
  {journal} {Phys. Rev. B}\ }\textbf {\bibinfo {volume} {30}},\ \bibinfo
  {pages} {6921} (\bibinfo {year} {1984})}\BibitemShut {NoStop}%
\bibitem [{\citenamefont {Lichtenstein}\ \emph {et~al.}(2001)\citenamefont
  {Lichtenstein}, \citenamefont {Katsnelson},\ and\ \citenamefont
  {Kotliar}}]{li.ka.01}%
  \BibitemOpen
  \bibfield  {author} {\bibinfo {author} {\bibfnamefont {A.~I.}\ \bibnamefont
  {Lichtenstein}}, \bibinfo {author} {\bibfnamefont {M.~I.}\ \bibnamefont
  {Katsnelson}},\ and\ \bibinfo {author} {\bibfnamefont {G.}~\bibnamefont
  {Kotliar}},\ }\bibfield  {title} {\bibinfo {title} {Finite-temperature
  magnetism of transition metals: An ab initio dynamical mean-field theory},\
  }\href {https://doi.org/10.1103/PhysRevLett.87.067205} {\bibfield  {journal}
  {\bibinfo  {journal} {Phys. Rev. Lett.}\ }\textbf {\bibinfo {volume} {87}},\
  \bibinfo {pages} {067205} (\bibinfo {year} {2001})}\BibitemShut {NoStop}%
\bibitem [{\citenamefont {Braun}\ \emph {et~al.}(2006)\citenamefont {Braun},
  \citenamefont {Min\'ar}, \citenamefont {Ebert}, \citenamefont {Katsnelson},\
  and\ \citenamefont {Lichtenstein}}]{br.mi.06}%
  \BibitemOpen
  \bibfield  {author} {\bibinfo {author} {\bibfnamefont {J.}~\bibnamefont
  {Braun}}, \bibinfo {author} {\bibfnamefont {J.}~\bibnamefont {Min\'ar}},
  \bibinfo {author} {\bibfnamefont {H.}~\bibnamefont {Ebert}}, \bibinfo
  {author} {\bibfnamefont {M.~I.}\ \bibnamefont {Katsnelson}},\ and\ \bibinfo
  {author} {\bibfnamefont {A.~I.}\ \bibnamefont {Lichtenstein}},\ }\bibfield
  {title} {\bibinfo {title} {Spectral function of ferromagnetic $3d$ metals: A
  self-consistent $\mathrm{LSDA}+\mathrm{DMFT}$ approach combined with the
  one-step model of photoemission},\ }\href
  {https://doi.org/10.1103/PhysRevLett.97.227601} {\bibfield  {journal}
  {\bibinfo  {journal} {Phys. Rev. Lett.}\ }\textbf {\bibinfo {volume} {97}},\
  \bibinfo {pages} {227601} (\bibinfo {year} {2006})}\BibitemShut {NoStop}%
\bibitem [{\citenamefont {Koloren\ifmmode~\check{c}\else \v{c}\fi{}}\ \emph
  {et~al.}(2012)\citenamefont {Koloren\ifmmode~\check{c}\else \v{c}\fi{}},
  \citenamefont {Poteryaev},\ and\ \citenamefont {Lichtenstein}}]{ko.po.12}%
  \BibitemOpen
  \bibfield  {author} {\bibinfo {author} {\bibfnamefont {J.}~\bibnamefont
  {Koloren\ifmmode~\check{c}\else \v{c}\fi{}}}, \bibinfo {author}
  {\bibfnamefont {A.~I.}\ \bibnamefont {Poteryaev}},\ and\ \bibinfo {author}
  {\bibfnamefont {A.~I.}\ \bibnamefont {Lichtenstein}},\ }\bibfield  {title}
  {\bibinfo {title} {Valence-band satellite in ferromagnetic nickel: Lda+dmft
  study with exact diagonalization},\ }\href
  {https://doi.org/10.1103/PhysRevB.85.235136} {\bibfield  {journal} {\bibinfo
  {journal} {Phys. Rev. B}\ }\textbf {\bibinfo {volume} {85}},\ \bibinfo
  {pages} {235136} (\bibinfo {year} {2012})}\BibitemShut {NoStop}%
\bibitem [{\citenamefont {Vidberg}\ and\ \citenamefont
  {Serene}(1977)}]{vi.se.77}%
  \BibitemOpen
  \bibfield  {author} {\bibinfo {author} {\bibfnamefont {H.~J.}\ \bibnamefont
  {Vidberg}}\ and\ \bibinfo {author} {\bibfnamefont {J.~W.}\ \bibnamefont
  {Serene}},\ }\bibfield  {title} {\bibinfo {title} {Solving the {E}liashberg
  equations by means of {N}-point {P}ad\'e approximants},\ }\href@noop {}
  {\bibfield  {journal} {\bibinfo  {journal} {J. Low Temp. Phys.}\ }\textbf
  {\bibinfo {volume} {29}},\ \bibinfo {pages} {179} (\bibinfo {year}
  {1977})}\BibitemShut {NoStop}%
\bibitem [{\citenamefont {Weh}\ \emph {et~al.}(2020)\citenamefont {Weh},
  \citenamefont {Otsuki}, \citenamefont {Schnait}, \citenamefont {Evertz},
  \citenamefont {Eckern}, \citenamefont {Lichtenstein},\ and\ \citenamefont
  {Chioncel}}]{we.ot.20}%
  \BibitemOpen
  \bibfield  {author} {\bibinfo {author} {\bibfnamefont {A.}~\bibnamefont
  {Weh}}, \bibinfo {author} {\bibfnamefont {J.}~\bibnamefont {Otsuki}},
  \bibinfo {author} {\bibfnamefont {H.}~\bibnamefont {Schnait}}, \bibinfo
  {author} {\bibfnamefont {H.~G.}\ \bibnamefont {Evertz}}, \bibinfo {author}
  {\bibfnamefont {U.}~\bibnamefont {Eckern}}, \bibinfo {author} {\bibfnamefont
  {A.~I.}\ \bibnamefont {Lichtenstein}},\ and\ \bibinfo {author} {\bibfnamefont
  {L.}~\bibnamefont {Chioncel}},\ }\bibfield  {title} {\bibinfo {title}
  {Spectral properties of heterostructures containing half-metallic
  ferromagnets in the presence of local many-body correlations},\ }\href
  {https://doi.org/10.1103/PhysRevResearch.2.043263} {\bibfield  {journal}
  {\bibinfo  {journal} {Phys. Rev. Res.}\ }\textbf {\bibinfo {volume} {2}},\
  \bibinfo {pages} {043263} (\bibinfo {year} {2020})}\BibitemShut {NoStop}%
\bibitem [{\citenamefont {Pines}\ and\ \citenamefont
  {Nozieres}(1966)}]{pi.no.66}%
  \BibitemOpen
  \bibfield  {author} {\bibinfo {author} {\bibfnamefont {D.}~\bibnamefont
  {Pines}}\ and\ \bibinfo {author} {\bibfnamefont {P.}~\bibnamefont
  {Nozieres}},\ }\href@noop {} {\emph {\bibinfo {title} {Quantum liquids}}}\
  (\bibinfo  {publisher} {Benjamin, New York},\ \bibinfo {year}
  {1966})\BibitemShut {NoStop}%
\bibitem [{\citenamefont {Negele}\ and\ \citenamefont
  {Orland}(1988)}]{ne.or.88}%
  \BibitemOpen
  \bibfield  {author} {\bibinfo {author} {\bibfnamefont {J.}~\bibnamefont
  {Negele}}\ and\ \bibinfo {author} {\bibfnamefont {H.}~\bibnamefont
  {Orland}},\ }\href@noop {} {\emph {\bibinfo {title} {Quantum Many-Body
  Systems}}}\ (\bibinfo  {publisher} {Frontiers in physics (Addison-Wesley,
  Redwood City, Calif. ua, 1988)},\ \bibinfo {year} {1988})\BibitemShut
  {NoStop}%
\bibitem [{\citenamefont {Gull}\ \emph {et~al.}(2011)\citenamefont {Gull},
  \citenamefont {Millis}, \citenamefont {Lichtenstein}, \citenamefont
  {Rubtsov}, \citenamefont {Troyer},\ and\ \citenamefont {Werner}}]{gu.mi.11}%
  \BibitemOpen
  \bibfield  {author} {\bibinfo {author} {\bibfnamefont {E.}~\bibnamefont
  {Gull}}, \bibinfo {author} {\bibfnamefont {A.~J.}\ \bibnamefont {Millis}},
  \bibinfo {author} {\bibfnamefont {A.~I.}\ \bibnamefont {Lichtenstein}},
  \bibinfo {author} {\bibfnamefont {A.~N.}\ \bibnamefont {Rubtsov}}, \bibinfo
  {author} {\bibfnamefont {M.}~\bibnamefont {Troyer}},\ and\ \bibinfo {author}
  {\bibfnamefont {P.}~\bibnamefont {Werner}},\ }\bibfield  {title} {\bibinfo
  {title} {Continuous-time monte carlo methods for quantum impurity models},\
  }\href {https://doi.org/10.1103/RevModPhys.83.349} {\bibfield  {journal}
  {\bibinfo  {journal} {Rev. Mod. Phys.}\ }\textbf {\bibinfo {volume} {83}},\
  \bibinfo {pages} {349} (\bibinfo {year} {2011})}\BibitemShut {NoStop}%
\bibitem [{\citenamefont {Vora}(2010)}]{vora.10}%
  \BibitemOpen
  \bibfield  {author} {\bibinfo {author} {\bibfnamefont {A.~M.}\ \bibnamefont
  {Vora}},\ }\bibfield  {title} {\bibinfo {title} {Superconducting state
  parameters of binary alloys},\ }\href
  {https://doi.org/10.1016/j.physc.2010.04.001} {\bibfield  {journal} {\bibinfo
   {journal} {Physica C: Superconductivity}\ }\textbf {\bibinfo {volume}
  {470}},\ \bibinfo {pages} {475} (\bibinfo {year} {2010})}\BibitemShut
  {NoStop}%
\bibitem [{\citenamefont {Sasaki}\ and\ \citenamefont {Muto}(1990)}]{sa.mu.90}%
  \BibitemOpen
  \bibfield  {author} {\bibinfo {author} {\bibfnamefont {T.}~\bibnamefont
  {Sasaki}}\ and\ \bibinfo {author} {\bibfnamefont {Y.}~\bibnamefont {Muto}},\
  }\bibfield  {title} {\bibinfo {title} {Low temperature resistivity anomaly in
  tinb and tiv alloys},\ }\href
  {https://doi.org/https://doi.org/10.1016/S0921-4526(90)80995-U} {\bibfield
  {journal} {\bibinfo  {journal} {Physica B: Condensed Matter}\ }\textbf
  {\bibinfo {volume} {165-166}},\ \bibinfo {pages} {291} (\bibinfo {year}
  {1990})},\ \bibinfo {note} {proceedings of the 19th International Conference
  on Low Temperature Physics}\BibitemShut {NoStop}%
\bibitem [{\citenamefont {{Prekul}}\ \emph
  {et~al.}(1975{\natexlab{a}})\citenamefont {{Prekul}}, \citenamefont
  {{Rassokhin}},\ and\ \citenamefont {{Volkenshte{\v{i}}n}}}]{pr.ra.75a}%
  \BibitemOpen
  \bibfield  {author} {\bibinfo {author} {\bibfnamefont {A.~F.}\ \bibnamefont
  {{Prekul}}}, \bibinfo {author} {\bibfnamefont {V.~A.}\ \bibnamefont
  {{Rassokhin}}},\ and\ \bibinfo {author} {\bibfnamefont {N.~V.}\ \bibnamefont
  {{Volkenshte{\v{i}}n}}},\ }\bibfield  {title} {\bibinfo {title} {{Effect of
  spin fluctuations on the superconducting and normal properties of alloys of
  Ti containing V, Nb, or Ta}},\ }\href@noop {} {\bibfield  {journal} {\bibinfo
   {journal} {Soviet Journal of Experimental and Theoretical Physics}\ }\textbf
  {\bibinfo {volume} {40}},\ \bibinfo {pages} {1134} (\bibinfo {year}
  {1975}{\natexlab{a}})}\BibitemShut {NoStop}%
\bibitem [{\citenamefont {{Prekul}}\ \emph
  {et~al.}(1975{\natexlab{b}})\citenamefont {{Prekul}}, \citenamefont
  {{Rassokhin}},\ and\ \citenamefont {{Volkenshte{\v{i}}n}}}]{pr.ra.75b}%
  \BibitemOpen
  \bibfield  {author} {\bibinfo {author} {\bibfnamefont {A.~F.}\ \bibnamefont
  {{Prekul}}}, \bibinfo {author} {\bibfnamefont {V.~A.}\ \bibnamefont
  {{Rassokhin}}},\ and\ \bibinfo {author} {\bibfnamefont {N.~V.}\ \bibnamefont
  {{Volkenshte{\v{i}}n}}},\ }\bibfield  {title} {\bibinfo {title} {{Cause of
  low-temperature resistance maximum in titanium alloys}},\ }\href@noop {}
  {\bibfield  {journal} {\bibinfo  {journal} {Soviet Journal of Experimental
  and Theoretical Physics Letters}\ }\textbf {\bibinfo {volume} {22}},\
  \bibinfo {pages} {209} (\bibinfo {year} {1975}{\natexlab{b}})}\BibitemShut
  {NoStop}%
\bibitem [{\citenamefont {Denton}\ and\ \citenamefont
  {Ashcroft}(1991)}]{de.as.91}%
  \BibitemOpen
  \bibfield  {author} {\bibinfo {author} {\bibfnamefont {A.~R.}\ \bibnamefont
  {Denton}}\ and\ \bibinfo {author} {\bibfnamefont {N.~W.}\ \bibnamefont
  {Ashcroft}},\ }\bibfield  {title} {\bibinfo {title} {Vegard's law},\ }\href
  {https://doi.org/10.1103/PhysRevA.43.3161} {\bibfield  {journal} {\bibinfo
  {journal} {Phys. Rev. A}\ }\textbf {\bibinfo {volume} {43}},\ \bibinfo
  {pages} {3161} (\bibinfo {year} {1991})}\BibitemShut {NoStop}%
\bibitem [{\citenamefont {Zimanyi}\ and\ \citenamefont
  {Abrahams}(1990)}]{zi.ab.90}%
  \BibitemOpen
  \bibfield  {author} {\bibinfo {author} {\bibfnamefont {G.~T.}\ \bibnamefont
  {Zimanyi}}\ and\ \bibinfo {author} {\bibfnamefont {E.}~\bibnamefont
  {Abrahams}},\ }\bibfield  {title} {\bibinfo {title} {Disorder and
  interactions in the hubbard model},\ }\href
  {https://doi.org/10.1103/PhysRevLett.64.2719} {\bibfield  {journal} {\bibinfo
   {journal} {Phys. Rev. Lett.}\ }\textbf {\bibinfo {volume} {64}},\ \bibinfo
  {pages} {2719} (\bibinfo {year} {1990})}\BibitemShut {NoStop}%
\bibitem [{\citenamefont {Dobrosavljevi\ifmmode~\acute{c}\else \'{c}\fi{}}\
  and\ \citenamefont {Kotliar}(1993)}]{do.ko.93}%
  \BibitemOpen
  \bibfield  {author} {\bibinfo {author} {\bibfnamefont {V.}~\bibnamefont
  {Dobrosavljevi\ifmmode~\acute{c}\else \'{c}\fi{}}}\ and\ \bibinfo {author}
  {\bibfnamefont {G.}~\bibnamefont {Kotliar}},\ }\bibfield  {title} {\bibinfo
  {title} {Hubbard models with random hopping in d=\ensuremath{\infty}},\
  }\href {https://doi.org/10.1103/PhysRevLett.71.3218} {\bibfield  {journal}
  {\bibinfo  {journal} {Phys. Rev. Lett.}\ }\textbf {\bibinfo {volume} {71}},\
  \bibinfo {pages} {3218} (\bibinfo {year} {1993})}\BibitemShut {NoStop}%
\bibitem [{\citenamefont {Dobrosavljevi\ifmmode~\acute{c}\else \'{c}\fi{}}\
  and\ \citenamefont {Kotliar}(1994)}]{do.ko.94}%
  \BibitemOpen
  \bibfield  {author} {\bibinfo {author} {\bibfnamefont {V.}~\bibnamefont
  {Dobrosavljevi\ifmmode~\acute{c}\else \'{c}\fi{}}}\ and\ \bibinfo {author}
  {\bibfnamefont {G.}~\bibnamefont {Kotliar}},\ }\bibfield  {title} {\bibinfo
  {title} {Strong correlations and disorder in d=\ensuremath{\infty} and
  beyond},\ }\href {https://doi.org/10.1103/PhysRevB.50.1430} {\bibfield
  {journal} {\bibinfo  {journal} {Phys. Rev. B}\ }\textbf {\bibinfo {volume}
  {50}},\ \bibinfo {pages} {1430} (\bibinfo {year} {1994})}\BibitemShut
  {NoStop}%
\bibitem [{\citenamefont {Clusius}\ \emph {et~al.}(1960)\citenamefont
  {Clusius}, \citenamefont {Franzosini},\ and\ \citenamefont
  {Piesbergen}}]{cl.fr.60}%
  \BibitemOpen
  \bibfield  {author} {\bibinfo {author} {\bibfnamefont {K.}~\bibnamefont
  {Clusius}}, \bibinfo {author} {\bibfnamefont {P.}~\bibnamefont
  {Franzosini}},\ and\ \bibinfo {author} {\bibfnamefont {U.}~\bibnamefont
  {Piesbergen}},\ }\bibfield  {title} {\bibinfo {title} {Ergebnisse der
  tieftemperaturforschung xxxii. die atom - und elektronenwärme des vanadins
  und niobs zwischen 10$^\circ$ und 273$^\circ$~k},\ }\href
  {https://doi.org/https://zfn.mpdl.mpg.de/data/Reihe_A/15/ZNA-1960-15a-0728.pdf}
  {\bibfield  {journal} {\bibinfo  {journal} {Z. Naturforsch.}\ }\textbf
  {\bibinfo {volume} {15 a}},\ \bibinfo {pages} {728} (\bibinfo {year}
  {1960})}\BibitemShut {NoStop}%
\bibitem [{\citenamefont {Allen}\ and\ \citenamefont {Dynes}(1975)}]{al.dy.75}%
  \BibitemOpen
  \bibfield  {author} {\bibinfo {author} {\bibfnamefont {P.~B.}\ \bibnamefont
  {Allen}}\ and\ \bibinfo {author} {\bibfnamefont {R.~C.}\ \bibnamefont
  {Dynes}},\ }\bibfield  {title} {\bibinfo {title} {Transition temperature of
  strong-coupled superconductors reanalyzed},\ }\href
  {https://doi.org/10.1103/PhysRevB.12.905} {\bibfield  {journal} {\bibinfo
  {journal} {Phys. Rev. B}\ }\textbf {\bibinfo {volume} {12}},\ \bibinfo
  {pages} {905} (\bibinfo {year} {1975})}\BibitemShut {NoStop}%
\end{thebibliography}%

\end{document}